\documentstyle[12pt]{article}                         

\newcommand{\beq}{\begin{equation}}
\newcommand{\eeq}{\end{equation}}
\newcommand{\bq}{\begin{quotation}}
\newcommand{\eq}{\end{quotation}}
\newcommand{\bc}{\begin{center}}
\newcommand{\ec}{\end{center}}
\newcommand{\BFACE}[1] {\mbox{\boldmath $#1$} }

\def\ss{\subsection*}
\def\sss{\subsubsection*}

\begin{document} 

\title{{\sc The Quantum Vacuum and the Cosmological Constant Problem}}
\author{
{\sc S.E. Rugh\thanks{Symposion, Helleb\ae kgade 27, Copenhagen N, Denmark
({\em e-mail: rugh@symposion.dk)}} and  H.  
Zinkernagel\thanks{Instituto de Filosof\'\i a, CSIC, 
Calle Pinar 25, Madrid ({\em e-mail: zink@ifs.csic.es})}}  }
\date{}
\maketitle

\begin{quotation}
\noindent
\small
{\bf Abstract} - The cosmological constant problem arises at the
intersection between general relativity and quantum field theory,
and is regarded as a fundamental problem in modern physics. In
this paper we describe the historical and conceptual origin of the
cosmological constant problem which is intimately connected to the
vacuum concept in quantum field theory. We critically discuss how
the problem rests on the notion of physical real vacuum energy,
and which relations between general relativity and quantum field
theory are assumed in order to make the problem well-defined.
\end{quotation} 

\normalsize

\begin{center}
\subsection*{Introduction}
\end{center}

\ \\ 
Is empty space really empty?  In the quantum field theories (QFT's)
which underlie modern particle physics, the notion of empty space has
been replaced with that of a vacuum state, defined to be the ground
(lowest energy density) state of a collection of quantum fields.  A
peculiar and truly quantum mechanical feature of the quantum fields is
that they exhibit zero-point fluctuations everywhere in space, even in
regions which are otherwise `empty' (i.e. devoid of matter and
radiation). These zero-point fluctuations of the quantum fields, as
well as other `vacuum phenomena' of quantum field theory, give rise to
an enormous vacuum energy density $\rho_{vac}$. As we shall see, this
vacuum energy density is believed to act as a contribution to the
cosmological constant $\Lambda$ appearing in Einstein's field
equations from 1917,
\begin{equation}
\label{Einstein1917} R_{\mu \nu} - \frac{1}{2} g_{\mu \nu} R -
\Lambda g_{\mu \nu} = \frac{8 \pi G}{c^4} \; T_{\mu \nu}
\end{equation}
where $R_{\mu \nu}$ and $R$ refer to the curvature of spacetime,
$g_{\mu \nu}$ is the metric, 
$T_{\mu \nu}$ the energy-momentum 
tensor, $G$ the gravitational constant, and $c$ the speed of light.
The constant $\kappa = 8\pi G/c^4$ is determined by the criterion
that the equations should correspond to Newtonian theory in the limit
for weak gravitational fields and small velocities, and this
correspondence also constrains the value of $\Lambda$.
In fact, confrontation of eq.(\ref{Einstein1917}) with observations 
shows that $\Lambda$ is very small: Solar system and galactic 
observations already put an upper bound on the
$\Lambda$-term, but the tightest bound come from large scale cosmology
(see e.g. Carroll et al (1992) \cite{Carrolletal92}), 
\beq
|\Lambda| < 10^{-56} cm^{-2}
\eeq
This bound is usually interpreted as a bound on the vacuum energy 
density in QFT\footnote{The unit $GeV^4$ for the
energy density is a consequence of the particle physics units
$\hbar=c=1$.}: 
\beq 
| \rho_{vac} | < 10^{-29} \; g/cm^3 \sim 10^{-47} \; GeV^4 \sim 10^{-9}
\; erg/cm^3
\label{obsbound}
\eeq
By contrast, theoretical estimates of various contributions to
the vacuum energy density in QFT exceed the observational bound
by at least 40 orders of magnitude. This large discrepancy
constitutes the cosmological constant problem. More generally, one can
distinguish at least three different meanings to the notion of a
cosmological constant problem:\\ 

\begin{enumerate}
\item {\em A `physics' problem: QFT vacuum  $\leftrightarrow$ $\Lambda$}. 
Various contributions to the vacuum energy density are estimated from the
quantum field theories which describe the known particles and
forces. The vacuum energy density associated with these theories is
believed to have experimentally demonstrated consequences and is
therefore taken to be physical real. The cosmological implications
of this vacuum energy density follows when certain assumptions are made about the
relation between general relativity and QFT.

\item {\em An `expected scale' problem for $\Lambda$}. 
Dimensional considerations of some future theory of quantum gravity
involving a fundamental scale -- e.g. the Planck scale -- lead physicists
to expect that the cosmological constant, as well as other dimensional
quantities, is of the order $\sim 1$ in Planck units (for example,
$\Lambda$ should be of the order of Planck energy
densities).\footnote{This way of formulating the problem resembles
other `hierarchy problems' such as the problem of why the masses in the
Standard Model of particle physics are so small relative to (Planck) 
scales of a presumed more fundamental theory.}

\item {\em An `astronomical' problem of observing $\Lambda$}.
Astronomers and cosmologists may refer to the `cosmological
constant problem' as a problem of whether a small cosmological
constant is needed to reconcile various cosmological models with 
observational data. 

\end{enumerate}

\noindent
Although we will indicate how these different notions of the
cosmological constant problem are related, we shall in this paper be
almost exclusively concerned with the first of these
formulations. Accordingly, when we refer to the term `cosmological
constant problem' we normally mean 1.

In this manuscript we critically discuss the origin of the QFT
vacuum concept (see also \cite{rugh96,rugh98}), and attempt to
provide a conceptual and historical clarification of the
cosmological constant problem. The paper is organized as follows:
We first trace the historical origin of the cosmological constant
problem in the QFT context. We then review the basis for various
contributions to the vacuum energy density, from quantum
electrodynamics, to the electroweak theory with spontaneous
symmetry breaking, to quantum chromodynamics, and discuss how
symmetry breakings are assumed to have changed the vacuum energy
density in the early universe. Third, we present some critical
remarks on the substantial conception of the vacuum in QFT.
Fourth, we indicate exactly how the energy density of the vacuum
state of QFT is assumed to be related to Einstein's cosmological
constant, and discuss whether the problem is well-defined in the
curved spacetime background of our universe. We then attempt to
classify the various solution types to the cosmological constant
problem. Finally, we discuss physicists opinions of the status of
the problem, and point to the inherent, partly philosophical,
assumptions associated with the conception of the cosmological
constant as constituting a serious problem for modern physics.

\begin{center}
\ss{The `quantum' history of $\Lambda$} 
\end{center}

The cosmological constant
has been in and out of Einstein's equations (\ref{Einstein1917})
ever since Einstein introduced it in 1917 in order to
counterbalance gravitation and thus secure a static universe
\cite{north65,kragh96,ray90}. At least four phases in this history
can be discerned (see e.g. \cite{north65,petrosian74,sahni99}): 
1) Hubble's discovery of the expanding universe
eventually lead Einstein to dismiss the cosmological constant in
1931. 2) Already in 1927 Lema\^\i tre incorporated the cosmological
constant in his non-static model of the universe. During the 1930s
similar models were discussed, primarily in connection with the
so-called age problem, but more precise measurements of the Hubble
constant (which is related to the age of the universe) subsequently
undermined this motivation for cosmological models with a non-zero
$\Lambda$.  3) In the late 1960s Petrosian, Salpeter and Szekeres once
again re-introduced the cosmological constant to explain some peculiar
observations of quasars indicating a non-conventional expansion
history of the universe, but later data accumulation on quasars
removed also this motivation. 4) Recently, observations of supernovae
have indicated that a non-zero cosmological constant in the
cosmological models is needed after all.  However, confirmation of
this result by independent methods would be valuable, not least since
the interpretation of the recent data are dependent on assumptions
about supernovae which are questioned by the involved investigators
(B. Schmidt, private communication).\footnote{In this list of
historical motivations to introduce a non-zero $\Lambda$, one might
also mention inflation. In particular, if one wants to reconcile the
observations indicating that the matter content in the universe is not
sufficient to make up a flat universe with inflation models (which
almost generically predicts a flat universe), one is forced to
introduce a cosmological constant, see also Earman and Moster\' \i n
(1999) for a critical discussion of inflation \cite{earman99}.}

There are also interesting philosophical arguments connected to
this history of introduction and re-introduction of the
cosmological constant in general relativity, see e.g.
\cite{north65} and \cite{ray90}.\footnote{A detailed account of
the observational and philosophical motivation for $\Lambda$ in
general relativity is in preparation by Earman \cite{earman00}.}
But since our interest here is in the connection between the
quantum vacuum and the cosmological constant we shall leave as a
separate issue the philosophical and observational motivations for
$\Lambda$ which are not concerned with quantum field
theory.\footnote{As will be indicated below, however, the
`observational history' of the cosmological constant has
influenced the speculations of the quantum vacuum, for instance
those of Lema\^ \i tre and Zel´dovich. We also note that a need
for a small astronomical dictated $\Lambda$ -- if confirmed --
would add a further constraint on possible cancellation mechanisms
for the huge vacuum energy in quantum theory.}

In this section we shall thus be concerned with the less well known
history of the cosmological constant as seen from the quantum (vacuum)
point of view.  We will review this history with particular emphasis
on the events which have transformed or reconceptualized the
cosmological constant problem, in order to clarify how it came 
to be seen as a fundamental problem for modern physics. \\

\sss{Early history}

Inspired by the new ideas of quantum theory and Planck's law for
the radiation from a black body, Nernst already in 1916
\cite{nernst16} puts forward the proposition that the vacuum is
not `empty' but is a medium filled with radiation
[`Licht\"{a}ther'] which contains a large amount of
energy.\footnote{Kragh mentions that Nernst is the first to
address the energy content of the vacuum (\cite{kragh96} p.153).
For a discussion of the vacuum concept before the advent of
quantum ideas see e.g. the review by Saunders and Brown
\cite{saunders91}} At absolute zero temperature the energy density
of this `light ether' at frequency $\nu$ grows as $\sim \nu^3$, so
that the total energy density becomes infinite. One way to remedy
this problem is to assume a fundamental cut-off of frequency
$\nu_0$, but Nernst notes that even if we just consider the
radiation in the vacuum to vibrate with frequencies up to, say,
$\nu \sim 10^{20} \; s^{-1}$, the total energy content in this
radiation per cubic centimetre will still be larger than $ U \sim
1.52 \times 10^{23} Erg $. As Nernst puts it 
\begin{quote}
``Die Menge der im Vakuum vorhandenen Nullpunktsenergie ist
also ganz gewaltig,..'' [the amount of available zero-point energy is
therefore quite enormous] (\cite{nernst16} p. 89).  
\end{quote}
Nevertheless
Nernst' ideas about the energy content of the vacuum were not
used for any cosmological thoughts (his interests were in chemistry),
but rather to put forward a model of the water molecule.

A more solid foundation for speculations on the energy density of
the vacuum became available with the early developments in quantum
electrodynamics (QED) in the mid-late 1920s \cite{schweber94}. In
QED the electromagnetic field is treated as a collection of
quantized harmonic oscillators, and contrary to a classical
harmonic oscillator -- which can be completely at rest and have
zero energy -- each quantized harmonic oscillator has a
non-vanishing `zero-point' energy. Enz and Thellung (\cite{enz60}
p.842) have pointed out that Pauli already in his early years
(mid-late 1920s) were concerned about the gravitational effects of
such zero-point energy. According to Enz and Thellung, Pauli in
fact made a calculation showing that if the gravitational effect
of the zero-point energies was taken into account (applying a
cut-off on the zero-point energies at the classical electron
radius)\footnote{We shall see below that some cut-off must be
imposed on the expression for the total zero-point energy, in
order for this to remain finite.} the radius of the world ``nicht
einmal bis zum Mond reichen w{\"u}rde'' [would not even reach to
the moon] (\cite{enz60} p.842).\footnote{C.P. Enz notes to us in
private communication that Pauli's discussion on the gravitational
effect of zero-point energies mainly took place at caf\' e
conversations with, among others, Otto Stern. We thank Prof. Enz
for discussions on this point.} Pauli's concern with the question
of zero-point energies is also mentioned in a recent article by
Straumann in which it is noted that Pauli was ``quite amused" by
his calculation \cite{straumann99}. Straumann rederives Pauli's
result by inserting the calculated energy density of the vacuum in
an equation relating the radius of curvature and the energy
density, $\rho \sim 1/a^2$ (derived from Einstein's equations
for a static dust filled universe). When the constants are
properly taken care of the result is that the radius of the
universe is about $31$ km -- indeed much less than the distance to
the moon! 

Nevertheless, as Enz and Thellung also point out, in Pauli's
extensive {\em Handbuch der Physik} article from 1933 on the
general principles of wave mechanics (including quantization of
the electromagnetic field), one finds only weak traces of these
discussions: Pauli notes that it is ``konsequenter" [more
consistent] from scratch to exclude a zero-point energy for each
degree of freedom as this energy, evidently from experience, does
not interact with the gravitational field (\cite{pauli33}
p.250).\footnote{Pauli also notes that such an energy is in
principle unmeasurable as it cannot be emitted, absorbed, or
scattered. This point, however, is not argued in any detail.} In
his discussion Pauli seems well aware that the zero-point energies
can indeed be avoided -- if gravity is ignored -- by rearranging
the operators in the Hamiltonian using what was later to be called
`normal ordering', a point to which we shall return in the
following section. In spite of his (unpublished) `caf\' e
calculation', Pauli's early worries do not seem to have had much
impact on the community of quantum physicists. 

Furthermore, it seems that the speculations of a huge vacuum
energy connected with the ideas of Dirac (with his hole theory
from 1930, see e.g. \cite{schweber94}) and also the final version
of QED constructed by Schwinger, Feynman, and others from the late
1940s, did not prompt any interest in the possible gravitational
consequences of these theories.\footnote{Presumably unrelated
to his early cosmological concerns, Pauli was in fact a strong
critic of the ``filling up'' of the vacuum not least in connection
with Dirac's hole theory, see \cite{weisskopf83}.} This should not
be surprising considering the preoccupation with divergence
problems which plagued higher order calculations in QED until the
late 1940s. Nevertheless, the cosmological constant problem was
not completely forgotten in this period, as evidenced in a 
quote from a conference address by Bohr in 1948:

\begin{quote}
...attention may also be called to the apparent paradoxes
involved in the quantum theory of fields as well as in Dirac's
electron theory, which imply the existence in free space of an
energy density and electric density, respectively, which [...]
would be far too great to conform to the basis of general
relativity theory. (\cite{bohr48}, p.222) 
\end{quote}
As a suggestion towards a solution to the huge vacuum energy density
problem, Bohr contemplates compensation mechanisms between
positive and negative zero-point field energies but remarks that
``[a]t present, it would seem futile to pursue such considerations
more closely...". 

In his historical survey, Weinberg writes \cite{Weinberg89}:
\begin{quote} 
Perhaps surprisingly, it was a long time before
particle physicists began seriously to worry about this problem
despite the demonstration in the Casimir effect of the reality of
zero point energies. Since the cosmological upper bound on $ | <
\rho > + \lambda / 8 \pi G | $ was vastly less than any value
expected from particle theory, most particle theorists simply
assumed that for some unknown reason this quantity was 
zero.\footnote{In
this quote, $\lambda$ corresponds to our $\Lambda$ (eqn
(\ref{Einstein1917})) and $\rho$ is the vacuum energy density.}
\end{quote} 
As Weinberg indicates, the Casimir effect (predicted by Casimir in
1948) is normally taken to add support to the view that the vacuum
zero-point energies are real, as the effect is interpreted as a
pressure exerted by the zero-point energies of empty space (we shall
return to the validity of this interpretation below). In connection to
Weinberg's remark we note that the interest in the Casimir effect, as
judged from citation indices, was very limited in the 50s and 60s
\cite{rugh98}.

While quantum physicists thus had other things to worry about, a
relation between the cosmological constant and vacuum energy was
noticed in cosmology, although the possible large energy
content of the vacuum from QFT does not seem to have played any
role in the discussions of the cosmological constant in the
cosmology literature
(despite that Eddington, for instance, pursued strongly the idea of
a unity between quantum mechanics and cosmology, see e.g.
\cite{kilmister94} and \cite{north65} p.85ff). As mentioned
earlier, the Belgian cosmologist G. Lema\^ \i tre constructed
a model of the universe with a cosmological constant in
1927, and in 1934 Lema\^ \i
tre commented on what such a constant could mean 
(\cite{lemaitre33}, p.12):

\begin{quote}
Everything happens as though the energy {\em in vacuo}
would be different from zero. In order that absolute motion,
i.e. motion relative to vacuum, may not be detected, we must associate
a pressure $p = - \rho c^2$ to the density of energy $\rho c^2$
of vacuum. This is essentially the 
meaning of the cosmological
constant $\lambda$ [$\Lambda$ in eqn (\ref{Einstein1917})] 
which corresponds to a negative density of
vacuum $\rho_0$ according to
$$\rho_0 = \frac{{\lambda} c^2}{4 \pi G} \sim 10^{-27} gr./cm.^3$$
\end{quote}
Lema\^ \i tre's constraint on the energy density of the vacuum is
a result of observational limits, only slightly less restrictive (two
orders of magnitude) than the constraint nowadays. 
Nevertheless, although Lema\^ \i tre
provides a physical interpretation of $\Lambda$, he does not point
to the quantum mechanical content of the
vacuum which occupied theoretical physicists at the time. 
But it is, in fact, not obvious
that Lema\^ \i tre in 1934 was unaware of the vacuum energy
arising in quantum field theory. For instance he discusses
Heisenberg uncertainty relations for the electromagnetic field in
a short article \cite{lemaitre33b} from 1933 in connection with
the by-then newly formulated quantum principles for the
electromagnetic field.

\sss{Recent history}

In his review, Weinberg indicates that the first published discussion
of the contribution of quantum fluctuations to the cosmological
constant was a 1967 paper by Zel'dovich
\cite{zeldovich67}.\footnote{As noted above, it was believed at the
time that a non-zero vacuum density was needed to account for
observations of quasars \cite{petrosian74}} Zel'dovich does not
address why the zero-point energies of the fields do not build up a
huge cosmological constant.  So he assumes, rather {\em ad hoc}, that
the zero-point energies, as well as higher order electromagnetic
corrections to this, are effectively cancelled to zero in the
theory. What is left are the higher order corrections
where gravity is involved, and the spirit of Zel'dovich's paper is
that this `left over' of vacuum energy, acting as a cosmological
constant, might explain the quasar observations. However, Zel'dovich's
estimate still gives a contribution to the cosmological constant which
is a factor $\sim 10^8$ too large relative to what was needed to
explain the quasar observations.\footnote{In the estimate of the
gravitational vacuum energy contribution, Zel'dovich considers a
proton as an example. The vacuum state of the proton field contains virtual
pairs of particles (virtual proton-antiproton pairs) with an effective
density $ n \sim 1 / \lambda^3 $ where $\lambda =
\hbar/ m c$ is the Compton wavelength of the proton (note, $\lambda$
does not refer to the cosmological constant in this expression). 
Zel'dovich then considers the gravitational
interaction energy of these virtual pairs which is $G m^2 /
\lambda$ for one pair, thus leading to a contribution to an
effective energy density in the vacuum of the order of magnitude: $
\rho_0 \sim G m^2/\lambda \times 1/\lambda^3 = G m^6 c^4 / \hbar^4 $.}

In a longer article \cite{zeldovich68} from the following year
Zel'dovich emphasizes that zero-point energies of particle
physics theories cannot be ignored when gravitation is taken into
account, and since he explicitly discusses the discrepancy
between estimates of vacuum energy and observations, he is
clearly pointing to a cosmological constant problem. In
\cite{zeldovich68}, Zel'dovich arrives at a QED zero-point energy
(his formula (IX.1) p.392) 
\beq 
\rho_{vac} \sim m \left (\frac{mc}{\hbar} \right ) ^3 \sim 10^{17} g/cm^3
\:\;\; , \:\;\; \Lambda = 10^{-10} cm^{-2}
\eeq
where $m$ (the ultra-violet cut-off) is taken equal to the proton
mass. Zel'dovich notes that since this estimate exceeds
observational bounds by 46 orders of magnitude it is clear that
``...such an estimate has nothing in common with reality''. 

Weinberg notes that the ``serious worry" about the vacuum energy
seems to date from the early and mid-1970s where it was realized
that the spontaneous symmetry breaking mechanism invoked in the
electroweak theory might have cosmological consequences
(\cite{Weinberg89} p.3).\footnote{As we shall explain below,
spontaneous symmetry breaking in the cosmological context refers
to a phase transition at a certain temperature by which a symmetry
of the vacuum state is broken.} While the authors who first
pointed out the connection between cosmology and spontaneous
symmetry breaking (Linde \cite{linde74}, Dreitlein
\cite{dreitlein74}, and Veltman \cite{veltman75}) did worry about
vacuum energy and the cosmological constant, they did not,
however, unambiguously express such worries in terms of a
cosmological constant {\em problem}: Linde notes that in
elementary particle theory without spontaneous symmetry breaking
the vacuum energy is determined only up to an arbitrary constant
and hence that ``...the `old' theories of elementary particles
have yielded no information whatever on the value of $\Lambda$''.
Obviously, if there is no information at all on the value of
$\Lambda$ from particle theories, then there is no worry of a
cosmological constant problem. According to Linde, however, the
{\em difference} of vacuum energy density before and after
symmetry breaking is well defined.\footnote{Linde takes
spontaneous symmetry breaking to imply that the vacuum energy
density depends on temperature, and hence on time in a hot Big
Bang universe where the temperature decreases with time, although
``almost the entire change [in vacuum energy density]" occurs at
the time of the symmetry breaking (see also below).} Although
Linde does not directly speak of a cosmological constant problem,
he estimates that, given the observational constraints, the vacuum
energy density must have changed roughly 50 orders of magnitude
from times before the spontaneous symmetry breaking until today
and concludes that this "makes speculations concerning a nonzero
value of $\Lambda$ in the present epoch more likely". Like Linde,
Dreitlein \cite{dreitlein74} does not directly discuss the
cosmological constant as a problem. In fact, by the assumption
that the vacuum energy density vanishes before the symmetry
breaking, Dreitlein suggests that the observational constraints on
the cosmological constant now can be used to put constraints on
the Higgs mass. Veltman \cite{veltman75}, on the other hand, takes
Linde's result to imply a radical discrepancy between
observational limits of the cosmological constant and the
theoretical estimate of vacuum energy from the model of
electroweak symmetry breaking -- thus stating clearly the
cosmological constant problem in the context of spontaneous
symmetry breaking. Moreover, by pointing out a problem in
Dreitlein's assumptions which suggests that even a small Higgs
mass would produce effects which are excluded experimentally,
Veltman rejects Dreitlein's attempt of reconciling the Higgs
mechanism with cosmological observations and concludes that
Linde's result ``undermines the credibility of the Higgs
mechanism". In this sense, Veltman in fact cancels the
cosmological constant problem arising from electroweak spontaneous
symmetry breaking by suggesting that the Higgs mechanism could be
plain wrong.

Following these discussions, Bludman and Ruderman (1977)
\cite{bludman77} argue that even though the vacuum energy density was
very large at the time of the symmetry breaking, it was nevertheless
negligible in comparison with the thermal energy density of
ultra-relativistic particles present at the time. They point out that
the effect of this thermal energy density is ``to smooth out entirely
any consequences" of a large vacuum energy density, and thus that
there is no hope of neither confirming nor refuting the spontaneous
symmetry breaking hypothesis by means of observational constraints on
the cosmological constant. In fact they conclude:
\begin{quote}
The small or zero value observed for the cosmological constant
may suggest some supersymmetry or new gauge-invariance principle
to be discovered in some future supergravity theory or may simply
be a fundamental constant. To this old problem (or pseudo-
problem), neither broken symmetry nor we have anything to add.
\cite{bludman77}
\end{quote}
The term `pseudo-problem' is explained by noting that since any
value of $\Lambda$ can be obtained from field theory by adding
suitable counter terms (see next section) ``...the observed value
[of $\Lambda$] is a problem only if one takes the attitude that it
should be derivable from other fundamental constants in particle
physics". This is not further elaborated but Bludman and Ruderman
have a reference to Zel'dovich's 1968 paper in which it is
suggested that a possible relation between $\Lambda$ and
fundamental constants in particle physics might be ``useful in the
construction of a genuine logically-consistent theory"
(\cite{zeldovich68} p.384). We take it, therefore, that Bludman
and Ruderman suggest that $\Lambda$ might not be derivable from a
more fundamental theory (incorporating both gravity and particle
physics), and that, in any case, spontaneous symmetry breaking
does not help to resolve the cosmological constant issue.

It follows from this discussion that the appearance of spontaneous
symmetry breaking did not by itself constitute a consensus that
the cosmological consequences of the vacuum energy density result
in a fundamental problem for modern physics. In any case, the
advent of inflationary cosmology in the early 1980s stimulated
further interest in vacuum energy with cosmological effects.
Indeed Guth \cite{guth81} acknowledges Bludman and Ruderman's
result but argues that, contrary to the electroweak phase
transition, the vacuum energy density during the spontaneous
symmetry breaking of a Grand Unified Theory (GUT) is {\em
larger} than the thermal energy density.\footnote{GUT refers to
the idea, first suggested in the mid-1970s, of a theory in
which the description of the electroweak force is unified with
that of the strong nuclear force.} According to Guth, it is this
vacuum energy density which drives the exponential expansion of
the universe. Guth mentions that (\cite{guth81}, note 11):

\begin{quote}
The reason $\Lambda$ is so small is of course one of the {\em deep
mysteries} of physics. The value of of $\Lambda$ is not determined
by the particle theory alone, but must be fixed by whatever theory
couples particles to quantum gravity. This appears to be a 
separate problem from the ones discussed in this paper, and I 
merely use the empirical fact that $\Lambda \simeq 0$.
[emphasis added]
\end{quote}

But given that the assumption of inflation specifically {\em
needs} a large early vacuum energy density in the early universe
to produce an (anti-)gravitational effect, inflation actually
emphasizes the cosmological constant problem rather than being
separate from it.\footnote{Although some later versions of
inflation do not link the origin of inflation with a specific
particle physics model, e.g. some version of GUT, they are still
based on an (unspecified) form of vacuum energy.} This may also be
indicated by the fact that Guth refers to the problem as a ``deep
mystery" of physics, in sharp contrast to e.g. Bludman and
Ruderman's notion of a ``pseudo-problem". To conclude this short
historical survey we note that, according to Witten
(\cite{witten97} p.279, \cite{witten00}) the cosmological constant
problem has been, and remains to be, an important obstacle for the
development of string theory (we return to this connection later).
This, together with the development of inflationary cosmology
since the early 1980s, have contributed to a recognition of the
importance of the cosmological constant problem. In the following
section we shall briefly survey how the various components of QFT
contribute to the vacuum energy density which is believed to
imply a huge cosmological constant.

\begin{center}
\ss{The origin of the QFT vacuum energy density}   
\end{center}

For more than two decades it has been customary for particle
physicists to assert that the `Standard Model' of elementary particle
physics is an essentially correct model of microphysics up to energies
of the order $\sim 100 GeV$.  According to the Standard Model, matter
is made up of leptons and quarks which are interacting through three
basic types of interactions: The electromagnetic, the weak and the
strong interactions.  Whereas the electromagnetic and weak forces are
unified in the electroweak theory (Glashow-Salam-Weinberg theory), the
theory of strong interactions, quantum chromodynamics (QCD), comprises
a sector of its own.  The Standard Model includes an additional
coupling of its constituents (fields) to Higgs fields which play a
crucial role both in constructing the electroweak theory, and in
generating masses of the Standard Model particles.

Below we shall discuss energy estimates for the ground state (the
vacuum state) of the Standard Model in terms of individual vacuum
contributions from each of its sectors studied in
isolation.\footnote{The vacuum state for the fields in the
Standard Model ought to be discussed as a whole. However, studying
the various sectors in isolation is a good starting point as one
does not expect these sectors to be too strongly coupled (i.e. it
is expected that the total vacuum energy of the complete model is
roughly a sum of the vacuum energy contributions of the individual
sectors).} Apart from the contributions described
below, the vacuum energy density receives contributions from any
quantum fields which may exist but remain to be discovered.

\ \\
\sss{QED}
Of the basic components in the Standard Model, the quantum theory
of electromagnetic interactions (QED) is both the simplest, the
first, and the most successful example of a working quantum field
theory.\footnote{It is possible to toy with even simpler quantum
field theories, such as a scalar field theory, which do not have
the complexities of a gauge theory and which can therefore be
utilized, as a calculational device, for illustrative purposes in
QFT.} 

We first recall that the systems studied in non-relativistic
quantum mechanics have a finite number of degrees of freedom where
spatial coordinates, momenta, and the energy (the Hamiltonian) are
represented by quantum operators which are subjected to a set of
commutation relations. A very simple quantum system -- important
for many applications -- is the quantum harmonic oscillator. The
ground state (the state with lowest energy) of the quantum
harmonic oscillator has a non-vanishing zero-point energy $E_0 =
1/2 \hbar \omega$ (where $\omega$ is the oscillation frequency of
the corresponding classical harmonic oscillator).\footnote{The
zero-point energy is present in Planck's famous radiation law, and
is also important for many physical phenomena in low temperature
physics. The zero-point energy in non-relativistic quantum
mechanics is responsible, for example, for Helium remaining a
fluid even at the lowest accessible temperatures.}

\ \\
{\em The free electromagnetic field}

In a classical field theory, like classical electromagnetism,
there are infinitely many degrees of freedom; the electric and
magnetic fields have values ${\BFACE{E}} ({\BFACE{x}}, t)$,
${\BFACE{B}} ({\BFACE{x}}, t)$ at each spacetime point. The
electromagnetic field is quantized by imposing a set of
(canonical) commutation relations on the components of the
electric and magnetic fields.\footnote{The commutation relations
between the field components (first derived by Jordan and Pauli in
1928) may be inferred from the commutation algebra of the creation
and annihilation operators in terms of which the quantized
electromagnetic field components are written, see also e.g. Heiter
(\cite{heitler54}, pp.76-87).} In the quantization procedure
the classical fields are replaced by quantum operators defined in
each spacetime point, and -- in order to build up a quantum
theory with the correct classical limit -- the Hamiltonian density
of the quantum theory is taken to be the same function of the
field operators $\hat{{\BFACE{E}}}$ and $\hat{{\BFACE{B}}}$ as the
energy density in the classical theory ${\cal H}$ = 1/2
$({{\BFACE{E}}}^2 + {{\BFACE{B}}}^2 ) $. 

The vacuum state $|0>$ of a quantum field theory like QED
is defined as the ground state of the 
theory. It turns out that in the ground state 
$< 0| \hat{{\BFACE{E}}} |0> = 0$,   
$< 0|\hat{{\BFACE{B}}} |0> = 0 $ whereas
$< 0|\hat{{\BFACE{E}}}^2 |0> \neq 0$ and   
$< 0|\hat{{\BFACE{B}}}^2 |0> \neq 0$.
These non-zero values of the vacuum expectation values for the squared
field operators are often referred to as quantum field
fluctuations but one should not think of them as fluctuations in time: 
Since the vacuum state is a (lowest) energy eigenstate of the free QED 
Hamiltonian, there is no time evolution of this vacuum state.  

The total zero-point energy of the QED theory can be expressed by
(see e.g. \cite{milonni94} p.364)
\begin{equation} \label{discretecont}
E \;=  \; <0| \hat{H} | 0> \; =\; 
\frac{1}{2}<0|\int d^3 x({\hat{\bf E}}^2 + {\hat{\bf
B}}^2)|0> \;
= \; \delta^{3} (0) \int d^3 k \frac{1}{2} \hbar \omega_{\bf k}
\end{equation}
where $\omega_{\bf k}$ and ${\bf k}$ refer to frequencies and
wave-numbers of a continuum of (plane-wave) modes. The energy in
this expression is strongly divergent since the expression
involves the product of two infinite (divergent) quantities. One
can render the integration finite by imposing an ultraviolet
frequency cut-off (some large $\omega_{max}=c|{\bf{k}}|_{max}$)
signifying up to which frequency range one believes the theory.
The infinite delta-function $\delta^3 (0)$ can be regularized in a
more formal way by introducing a box of volume $V$.\footnote{The
``box regularization" (enclosing the field in a box with volume
$V$) implies the following replacement in the right hand side of
(\ref{discretecont}): $ \delta^3 (\BFACE{k}) = (1/2\pi)^3 \int d^3
x \; e^{i \BFACE{k} \BFACE{x}} \rightarrow V/8 \pi^3 $ for
$\BFACE{k} \rightarrow 0$, see e.g. (\cite{milonni94} p. 364.)}
The introduction of this regularizing $V$ resembles closely (in
the limit $V \rightarrow \infty$) the standard `box-quantization'
procedure for the electromagnetic field in which an artificial
`quantization volume' $V$ is used to exploit the formal
equivalence of a field mode with a harmonic quantum oscillator.
Once the representation of the electromagnetic field as a set of
harmonic oscillators is introduced, a finite expression for the
energy density (= energy per volume) can be derived directly from
the summed zero-point energy for each oscillator
mode,\footnote{When we here and in the following use the symbol
$\rho$ for vacuum energy densities we understand the quantum
mechanical expectation value $<0|\hat{\rho}|0>=<\rho_{vac}>$ for
the energy density operator in the vacuum state.}

\beq \label{zpeqed}
\rho_{vac} = \frac{E}{V} = 
\frac{1}{V} \sum_{\bf k} \frac{1}{2} \hbar \omega_{\bf k} 
\; \; \approx \; \; \frac{\hbar}{2 \pi^2 c^3}
 \int _{0}^{\omega_{max}} \omega ^3 d\omega
\; \; = \frac{\hbar}{8 \pi^2 c^3}  \omega _{max}^4 
\eeq
where the wave vector ${\bf k}$ now refers to the so-called normal
modes (of the electromagnetic field) which are compatible with the
boundary conditions provided by the box volume $V$.\footnote{Think
of a string fixed at two endpoints which will be able to vibrate
in certain `normal modes'. An electromagnetic field with `fixed'
values at the boundary of an `artificial box' will similarly be
able to vibrate in simple harmonic patterns (normal modes). The
expression (\ref{zpeqed}) is the summed energy of these field
modes with an upper cut-off $\omega_{max}$ setting an interval of
frequencies $0 \leq \omega \leq \omega_{max}$ in which the
description is viable (For the vibrating string, for example, the
minimal wavelength of the vibrations cannot be less than the
distance between the atoms).} The right hand side of the equation
follows from the left hand side in the limit $V \rightarrow
\infty$ where the energy density does not depend on the `box
quantization' volume $V$.\footnote{Thus, one can shift between the
sum and the integral (and thereby also obtain this equation from
eq.(\ref{discretecont})) by using the standard replacement formula
$\sum_{\bf{k}} \approx \frac{V}{8\pi^3} \int d^3 k$ which is a
good approximation in the limit $V \rightarrow \infty$. Note that
if there are real physical constraints on the normal modes ${\bf
k}$, for example provided by the physical boundaries of the
Casimir plates involved in the Casimir effect, such a replacement
(such an approximation) of a sum by an integral is
inappropriate. In fact, such a replacement will `approximate away'
the Casimir effect since the Casimir energy between such
boundaries can be traced to the difference between a discrete sum
and a continuous integral, see also e.g. (\cite{milonni94} p.57).}

Before providing some order-of-magnitude
estimates for vacuum energy densities, corresponding to 
different ultraviolet frequency cut-off's in expression (\ref{zpeqed}), 
we shall briefly discuss what happens when interactions 
are taken into account. 

\ \\
{\em Interacting electromagnetic fields}

The zero-point energy discussed above is a lowest order
consequence of QED, i.e. it is present before interactions are
considered. To perform detailed calculations in QED, the
interactions are treated as small perturbations to the
non-interacting theory in powers of the so-called fine structure
constant $\alpha = 1/137$ (which determines the strength of the
interactions). When the coupling between the electromagnetic field
and the electron-positron fields is included, one often speaks of
the production and annihilation of virtual electron-positron pairs
in the `interacting' vacuum.\footnote{This popular picture is
actually misleading as no production or annihilation takes place
in the vacuum. The point is rather that, in the ground state of
the full interacting field system, it is not well-defined to speak
of a definite number of quanta (particles) for any of the fields.
For instance, the photon number operator does not commute with the
Hamiltonian for the interacting field system, hence one cannot
speak of a definte number (e.g. zero) of photons in the vacuum of
the full interacting system, see also e.g. (\cite{aitchison85}
p.353).} This `vacuum' of virtual particles, resulting from higher
order diagrams in QED, contributes further to the vacuum energy
density $\rho_{vac}$.\footnote{In perturbative QED, one would
expect these higher order contributions to the zero-point energy
to be suppressed by factors of $\alpha$, $\alpha ^2$, etc.
relative to the lowest order zero-point energy.} In standard QED
calculations these higher order contributions to the vacuum energy
-- so-called vacuum blob diagrams (without external lines) -- can
be ignored as they do not contribute to the scattering amplitudes
of physical processes (see e.g. \cite{berestetskii82} p.460).

In order to characterize the resulting picture of the interacting
vacuum, it is sometimes pointed out that a system of interacting
quantum fields is analogous to a complicated interacting quantum
mechanical system in solid state physics. For instance, like a
system in solid state physics, the system of interacting fields
can exist in different energy states, namely the ground state and
various exited states. The exited states of the fields system are
characterized by the presence of excitation quanta, which,
according to QFT, are the particles (electrons, quarks,
photons...) of which our material world is composed (see also
\cite{aitchison85}). As we will briefly discuss below, however,
there is an important difference between zero-point fluctuations
in physical (quantum mechanical) systems and zero-point
fluctuations of the interacting QFT vacuum. For whereas photons
(e.g. X-rays) are scattered on zero-point fluctuations in a
crystal lattice of atoms even when $T \rightarrow 0$, photons (or
anything else) do not scatter on the vacuum fluctuations in QED.
Indeed, if photons (light) were scattered on the vacuum
fluctuations in large amounts, astronomy based on the
observation of electromagnetic light from distant astrophysical
objects would be impossible. There is thus a break-down of the
analogy between the QED vacuum and the ground state of a quantum
mechanical system (e.g. in solid state physics).

\ \\
{\em Estimates of the QED zero-point energy}

How large is the zero-point energy in empty space supposed to be? The
numerical answer clearly depends on which frequency interval we
employ in the integration in eq.(\ref{zpeqed}). For field modes in
the optical region from 400 nm to 700 nm (visible light), the
corresponding zero-point energy will amount to about $220 \;
erg/cm^3$ (\cite{milonni94}, p. 49). If we instead consider the
electromagnetic field modes in the energy range from zero up to an
ultraviolet cut-off set by the electroweak scale $\sim 100 \; GeV$
(where the electromagnetic interaction is believed to be
effectively unified with the weak forces in the more general
framework of the electroweak interaction), a rough estimate of the
zero-point energy will be\footnote{We neglect factors like $8
\pi^2$ in equation (\ref{zpeqed}).}
$$
\rho^{EW}_{vac} \sim (100 \; GeV)^4 \sim 
10^{46} \; erg/cm^3
$$
This is already a huge amount of vacuum energy attributed to the
QED ground state which exceeds the observational bound
(\ref{obsbound}) on the total vacuum energy density in QFT by
$\sim 55$ orders of magnitude. Such an estimate is therefore more
than sufficient to establish a significant discrepancy between a
theoretically estimated vacuum energy and the observed
cosmological constant. To extrapolate quantum field theories
substantially above energies of the electroweak scale of $100 \;
GeV$ involves a strong element of speculation beyond what has been
tested experimentally, but it has nevertheless been customary to
imagine that the QFT framework is effectively valid up to scales
set by the Planck energy
$$
E_P = \left(\frac{\hbar c^5}{G} \right)^{1/2} \sim 10^{19} GeV
$$
It is easy to envisage, of course, that new physics can enter
between electroweak scales and Planck scales but if the
alternations are still within the framework of quantum field
theory, there will in general still be vacuum energy.\footnote{An
exception to this is e.g. supersymmetric QFT -- see below.}
Assuming this energy to be of the QED zero-point energy type,
we get roughly (by inserting the Planck energy in eq.(\ref{zpeqed}) 
with $E_P = \hbar \omega_{max}$), 
$$
\rho^{Planck}_{vac} \sim (10^{19} \; GeV)^4 \sim 10^{76} \; GeV^4
\sim 10^{114} \; erg/cm^3
$$
thus over-estimating the vacuum energy, relative to the 
observational constraint (\ref{obsbound}),
by more than $\sim 120$ orders of magnitude!

\sss{Electroweak theory and spontaneous symmetry breaking}

So far we have only discussed QED. When also the weak
interactions are considered -- responsible, for instance, for
radioactive $\beta$-decays -- the standard framework is the
electroweak theory. In order for this theory to describe massive
fermions and bosons (and remain renormalizable), one needs to
introduce a so-called Higgs field which gives masses to the
particles by means of `spontaneous symmetry breaking'. Generally,
spontaneous symmetry breaking refers to a situation where the
governing equations for the dynamics of the fields (the Lagrangian
of the theory) have a symmetry which is not shared by the vacuum
state -- the vacuum state breaks the symmetry in question. All the
massive particles in the Standard Model are coupled to the Higgs
field (via so-called Yukawa couplings) and their masses are
proportional to the vacuum expectation value of the Higgs field
which is non-zero in the broken phase (see e.g. Weinberg
\cite{weinberg96} and Brown and Cao \cite{brown91} for
respectively a physics textbook and a historical account of
spontaneous symmetry breaking).

Contrary to the other parts of the Standard Model, there remains
a considerable degree of choice for how to construct the Higgs 
sector.\footnote{It is well known that no Higgs particles have been
observed, and although the Higgs mechanism is the widely accepted 
way of giving masses to particles in the Standard Model, it is 
interesting that some physicists often consider the Higgs sector to be 
most unwanted, and encourage the search for alternative mechanisms
(these physicists include Veltman \cite{veltman86} and Glashow (talk
given at Les Houches (1991): `Particle Physics in the nineties').} 
For example, one may construct a Standard Model with
only one (complex) field $\phi$ (this is the case
in the simplest Standard Model), or there may be two, three, or
more Higgs fields.\footnote{In the following we consider the
simplest Higgs model. A motivation for contemplating a more
complicated architecture for the Higgs sector is the result -- as
revealed by numerical simulations -- that the Standard Model with a too
simple Higgs sector is insufficient to account for a surplus of
baryons over anti-baryons (baryon asymmetry) which is believed
to be a feature of our universe, see e.g.  Trodden \cite{trodden98}.}
The vacuum energy density resulting from the Higgs field is calculated
from the electroweak theory by noting that the scalar Higgs field
potential is of the form (see e.g. \cite{Weinberg89})

\begin{equation} \label{Higgspotential}
V(\phi) = V_0 - \mu ^2 \phi ^2 + g\phi ^4
\end{equation}
where $g$ is a Higgs (self) coupling constant and $\mu$ an energy
scale which is related to the vacuum expectation value $v$ of the
Higgs field ($\mu ^4 = g^2 v^4$).\footnote{The form of this potential
is not arbitrary: The requirements of $V(\phi)$ (and the Lagrangian
$\cal{L}$) to be symmetric under reflection ($\phi \leftrightarrow -
\phi$) and renormalizability of $\cal{L}$, constrain
$V(\phi)$ to be a polynomium with $\phi ^n$ terms up to at most fourth
order in the Higgs field $\phi$ (when the spacetime dimension is
four). I.e.  in spacetime dimension four, only a constant, and
$\phi^2$ and $\phi^4$ terms are possible.} The vacuum expectation
value of the Higgs field $v = < \phi >$ is inferred from the
experimentally known Fermi coupling constant (which in turn is
determined from the muon decay rate) to be $v = < \phi > \simeq 250 \;
GeV$.\footnote{The association of the vacuum expectation value of
the Higgs field with the Fermi coupling constant is due to the
identification of a electroweak interaction diagram (involving a
W-boson) with that of an effective four-Fermi coupling from the
phenomenologically successful `V-A' theory -- the predecessor to
electroweak theory, see e.g. \cite{weinberg96}, Vol. II, p.310.
Note that this non-zero vacuum expectation value for the Higgs
field implies that there should be real (not just virtual) Higgs
particles everywhere in space (in contrast to QED where
the vacuum state is a ``no photon'' state and).} The Higgs potential
(\ref{Higgspotential}) is minimized for 
$\phi ^2 = \mu ^4 / 4g$ where $V(\phi)$  takes the value 
$ V_{min} = V_0 - \mu ^4/4g (\; = \; \rho _{Higgs,vac})$.
If one (somewhat arbitrarily) assumes that $V(\phi)$ vanishes 
for $\phi = 0$, and if the 
electromagnetic fine structure constant
squared is taken as a reasonable estimate for the Higgs coupling constant 
$g$,\footnote{Weinberg quotes
this as a low estimate of $g$. Since the Higgs mass to lowest order in
the perturbative series is given by $m_{\phi} = gv$, a too low value
of $g$ would be inconsistent with the fact that the Higgs boson has
still not been observed.} we
are left with a Higgs vacuum energy density of the order of
$$
\rho _{Higgs,vac}= -\mu ^4/ 4g = -gv^4 \; \approx -10^5 \; 
{\mbox{GeV}} ^4 \; -10^{43} erg/cm^3
$$
which, in absolute value, is roughly 52 orders of magnitude larger than
the experimental bound on $\Lambda$ (quoted in eqn.
(\ref{obsbound})).

Like the other estimates of vacuum energy, this too is model
dependent. There are, for example, no convincing reasons to take
$V(\phi)=0$ for $\phi =0$, and one might as well assume that the
Higgs vacuum energy (including higher-order corrections) could be
cancelled by $V_0$. But, in any case, one would need an extreme
fine tuning to bring $\rho _{Higgs,vac}$ in accordance with the
observational bound on $\Lambda$. Moreover, finite temperature
quantum field theory applied to the electroweak theory of the
Standard Model gives the result that the Higgs field potential for
non-zero temperatures has correction terms which depend on the
temperature $T$ \cite{kolb93}. The lowest order correction term is
of the form $\sim T^2\phi ^2$ which, for sufficiently high $T$,
makes the potential take its minimum ($V=V_0$) at $\phi =0$. Thus,
even if one uses $V_0$ ($\simeq + 10^5 \mbox{GeV}^4$) to cancel
the Higgs field induced cosmological constant now, it must have
been very large at times before the electroweak phase transitions
(see also below).\footnote{Within the standard hot big bang theory
of the cosmos, this electroweak phase transition is believed to
have taken place approximately $10^{-11}$ seconds after the `Big
Bang'. The exact way in which the transition takes place from the high
temperature symmetric phase with $<\phi> = 0$ (at $T>T_c$) to the
low temperature asymmetric phase with $<\phi> = v \neq 0$ (at
$T<T_c$) can generally only be exploited by performing
rather involved numerical calculations. A tentative result is that
the phase transition among other things depends strongly on the
mass(es) of the Higgs field(s), see also e.g. (\cite{kolb93},
chapt. 7-8).} 

\sss{QCD}

Quantum chromo dynamics (QCD) is a theory which describes the
so-called strong interactions of quarks and gluons, the latter
representing the forces which e.g. bind together the constituents of
the nucleus.  In the low energy regime QCD is a non-perturbative and
highly non-linear theory, and thus its quantum states, in particular
its ground state, cannot with good approximation be expressed in terms
of harmonic oscillators.\footnote{In contrast to electromagnetic
interactions with a small coupling constant $\alpha = e^2 / \hbar c
\approx 1/137$, the coupling constant for strong interactions is large
in the low energy regime.  This makes the (low energy) strong coupling
constant essentially unfit as a perturbation parameter.}  This point
makes discussions of the QCD vacuum highly complicated and, although a
number of different models for the QCD vacuum has been developed (see
e.g. \cite{shuryak88}), the vacuum structure of QCD is far from being
a settled theoretical issue. Nevertheless, it is generally asserted
that the non-perturbative sector of QCD gives rise to gluon and quark
`condensates' in the vacuum at low energies (at zero temperatures),
that is, non-vanishing vacuum expectation values of the quark and
gluon fields, e.g.  $<0 | \bar{q} q| 0> \neq 0$.  Estimates of the
vacuum energy density associated with these condensates are rather
model dependent but they generally lead to vacuum energy densities
given by some pre-factor times $\lambda_{QCD}^4$. The quantity
$\lambda_{QCD} \sim 0.2-0.3 \; GeV$ is a characteristic scale for QCD
where the strong coupling constant is of order unity (separating the
perturbative and the non-perturbative regime).  One thus frequently
estimates:
$$
\rho_{QCD} \sim 10^{-3}-10^{-2} GeV^4 \sim 10^{35}-10^{36} erg/cm^3 
$$
which is more than 40 orders of magnitude larger than the observational
bound (\ref{obsbound}) on the total vacuum energy density. 

The scale $\lambda_{QCD}$ is also believed to set a temperature scale
marking a QCD ``phase transition" in which the quark and gluon
condensates in vacuum, present at lower temperatures, disappear as the
temperature increases. This phase transition is related to the
restoration of so-called chiral symmetry.  The picture is roughly as
follows: At low temperatures, the chiral symmetry is spontaneously
broken as mirrored in a non-vanishing quark condensate $<0 |\bar{q} q
| 0 > \neq 0$, whereas the chiral symmetry is restored above the phase
transition temperature where the quark condensate disappears, $< 0 |
\bar{q} q | 0 > = 0$. In addition, the high temperature phase is
characterized by the so-called quark-gluon plasma where the quarks are
no longer confined within the hadrons (deconfinement).
The experimental evidence for this picture of the QCD vacuum is,
however, not yet compelling.  For instance, there are -- so far -- no
clean cut experimental signatures of the QCD phase transition, and the
connection between the quark-gluon plasma to the chiral symmetry
breaking remains an interesting theoretical conjecture.\footnote{CERN
experiments are at present looking for signatures of the quark-gluon
plasma. The signals reported so far (so-called $J/\psi$ suppression,
etc.) are not considered unambiguous pointers to the existence of a
quark-gluon plasma or restoration of chiral symmetry. In fact, it is
far from clear that it is possible -- in a laboratory framework -- to
investigate QCD phase transitions which are of a duration of the order
of $10^{-23}$ seconds.}

As concerns the low temperature phase, the various models of the
QCD vacuum are constrained by experimental results, for instance
studies of the so-called charmonium decay (see e.g. Shuryak
\cite{shuryak88}, p.199). Moreover, theoretical analysis of the
chiral symmetry (chiral perturbation theory) is useful in
successfully predicting results of hadronic scattering experiments.
But the strongest evidence for the picture of the QCD vacuum is
often taken to be the observed properties of the pion. In
particular, the pion is observed to have a relatively small mass
which is in conformity with the picture of a spontaneous breakdown
of chiral symmetry.\footnote{Some physicists would find the
smallness of the pion mass a mystery to be explained by the fact
that the pion is, at first, considered to be a massless Goldstone
boson associated with the spontaneous breakdown of chiral
symmetry, and then its mass results, as a small correction, from
the pion's coupling to the non-vanishing quark condensate (see
below). However the ``hierarchy problem'' constituting this
mystery is hardly impressive: The pion is a factor $\; \approx 6
\approx 2 \pi \; $ less massive than other characteristic mass
scales, such as the mass of the proton etc., to which it is
compared.} The pion $\pi$ is comprised of an up ($u$) and a down
($d$) quark ``glued together'' by the strong interactions (with
highly complicated and non-perturbative dynamics)\footnote{See
e.g. S.E. Rugh, ``Chaos in Yang-Mills fields and the Einstein
equations'', Ph.D. thesis, The Niels Bohr Institute (1994).}. The
massless pion acquires a mass by virtue of the non-vanishing quark
condensate $< 0 | \bar{q} q | 0 > \neq 0$ in the vacuum:
\begin{equation} \label{GOR}
m_{\pi}^2 = - \; f_{\pi}^{-2} \; (\frac{m_u + m_d}{2})
< 0 | \bar{u} u + \bar{d} d | 0 >
\end{equation}
where $f_{\pi}$ is a constant related to the decay time of the
pion and $m_{u}$, $m_{d}$ is the mass of the $u$ and the $d$
quark, respectively. This `Gell-Mann-Oakes-Renner' relation
\cite{gellmann68,koch97}, connects a non-vanishing quark
condensate in the QCD vacuum with the pion mass.\footnote{The
relation is derived as a perturbative expansion in the small $u$
and $d$ quark masses anticipating -- in spirit -- what was later
more systematically developed as chiral perturbation theory, see
e.g. \cite{leutwyler94}. Similar relations can be put forward,
e.g. for the kaons $K^{0}, K^{+}, K^{-}$ which are related to a
condensate of s-quarks. But the s-quarks are much heavier, so
perturbative expansions in the s-quark masses are less clean.}
However, for such a relation to have predictive power it is
important that there are independent means of determining the $u$
and $d$ quark masses -- and these are hard to find. In fact, the
situation is rather the reverse since the relation (\ref{GOR}) is
employed to fit the $u$ and $d$ quark masses from the
experimentally determined value of the pion mass
\cite{weinberg77}! Moreover, is not clear that relation
(\ref{GOR}) points to properties of empty space since the pion is
considered to be build up of $u$ and $d$ quarks (`real'
excitations of quark degrees of freedom, not merely virtual quarks
associated with the quark condensate vacuum) including all the
`QCD glue' (the strong interactions) binding this system together.
As we will discuss below, the situation is in this respect
somewhat analogous to the Lamb shift in QED which is often taken
to point to the reality of vacuum fluctuations in empty space.
Also in the Lamb shift the `vacuum effect' is associated with a
real particle (an atom) comprised of a proton, an electron, and
`QED glue' (the radiation field). Hardly a direct pointer to empty
space.

Of course, the question of the QCD vacuum is very complicated. The
brief remarks offered here could, however, serve as pointers to
further scrutinization of the QCD conception of `empty
space'.\footnote{One could speculate if the idea of spontaneous
symmetry breakdown has to be a `global' vacuum effect of infinite
extension? Or could it be that, for instance, the $ \bar{q} q $
system (the $\pi$ meson) comprises a little finite volume system
with a spontaneously broken chiral symmetry -- just like a
ferromagnet can form a system with an apparent spontaneous
breakdown of (rotational) symmetry which is confined within a
finite volume?}

\sss{Phase transitions in the early universe}

Within the Big Bang framework it is assumed that the universe
expands and cools. During this cooling process the universe passes
through some critical temperatures corresponding to characteristic
energy scales of phase transitions. These transitions are
connected with symmetry breakings, each of which leaves the vacuum
state of the quantum fields less symmetric than 
before.\footnote{In finite temperature QFT, the ground state of the
quantum field system at a certain temperature is {\em not} a state without
excitation quanta (the temperature is associated with
the statistical properties of these particles).} A general picture thus
emerges of a more symmetric vacuum state (in earlier phases of the
universe) successively undergoing a chain of symmetry breakings
producing a less symmetric vacuum state at present. The physics of
the phase transitions, as well as the more exact position of the
critical temperatures of these transitions, depend of course on
the particle physics theory which is implemented to model the
content of the universe. An often envisaged example of a chain of
symmetry breakings could be illustrated as follows: 
$$ \cdots
\begin{array}{c} \\ \longrightarrow\\ _{\sim 10^{14} GeV}
\end{array} \left( \begin{array}{c} \mbox{GUT} \\
\mbox{symmetry}\\ \mbox{breaking} \end{array} \right)
\begin{array}{c} \\ \longrightarrow\\ _{\sim 10^{2} GeV}
\end{array} \left( \begin{array}{c} \mbox{Electroweak} \\
\mbox{symmetry}\\ \mbox{breaking} \end{array} \right)
\begin{array}{c} \\ \longrightarrow\\ _{\sim 10^{-1} GeV}
\end{array} \left( \begin{array}{c} \mbox{QCD} \\ \mbox{symmetry}
\\ \mbox{breaking} \end{array} \right) \cdots 
$$
The first of the symmetry breakings listed represents the phase
transition associated with the grand unified symmetry group which
breaks down to the symmetry group of the Standard
Model.\footnote{Note that one has very little, if any,
observational/experimental evidence constraining theory above
energy scales of $\sim 10^2$ GeV, so that phase transitions at
such energy scales remain purely speculative.} A characteristic
energy scale of grand unified symmetry breaking (see e.g.
\cite{kolb93}) is $\sim 10^{14} \; GeV$ (depending on model
assumptions), the characteristic energy of electroweak symmetry
breaking is $\sim 10^2 \; GeV$, and the characteristic energy of
chiral symmetry breaking in QCD is $\sim 0.1 \; GeV$. The
indicated (approximate) energy scales of the symmetry breakings
translate into order-of-magnitude expectations for associated
differences (between before and after the symmetry breaking) in
vacuum energy densities as, respectively, $\sim (10^{14} GeV)^4$,
$\sim (10^2 GeV)^4$, and $\sim (10^{-1} GeV)^{4}$.\footnote{In our
units of $\hbar = c = 1$, a characteristic energy of $E$ ($Gev$)
translates into a characteristic energy density of $E^4$
($GeV^4$).}

Under the assumption that vacuum energy density can be identified
with a cosmological constant, this implies a hierarchy of different
cosmological constants, one for each phase of the vacuum
state.\footnote{ The combination of counter terms (like a $V_0$ to
cancel a Higgs energy vacuum, cf. the previous section) -- needed to
cancel out the resulting vacuum energy density appearing today --
will get increasingly complicated as we take into account the
summed effect of all the symmetry breakings mentioned (A further
complication in achieving a zero for the vacuum energy density now
is that the quoted energy density, e.g. for the electroweak Higgs
potential is only the lowest order contribution, see also Coleman
in \cite{cao99} p.385). } Whereas we have a very tight
observational constraint (\ref{obsbound}) on the value of the
cosmological constant $\Lambda$ at present, there are only very
weak observational constraints on the effective vacuum energy
density (the effective cosmological constant) at earlier phases.
For example, if there was an effective vacuum energy density of
$\sim (10^2 GeV)^4$ before the electroweak phase transition, we
recall that Bludman and Ruderman \cite{bludman77} showed that this
(large) vacuum energy is negligible compared to characteristic
thermal energy densities of the particles at these early times. 

Since inflation is driven by vacuum energy, a large value of the
vacuum energy during the GUT phase transition is necessary from
the viewpoint of inflation models (at least, this was the original
idea of Guth in 1981 when he proposed the inflationary model). The
use of spontaneous symmetry breaking to account for inflation
requires a positive cosmological constant, which can be achieved
by the {\em ad hoc} assignment of a positive $V_0$ (cf.
eq.(\ref{Higgspotential})) to cancel the negative vacuum energy
density now. It is important to bear in mind, however, that while
inflation is held to solve various problems in the standard Big
Bang cosmology (the monopole, flatness and horizon problem, see
e.g. \cite{earman99}), it offers no clue to the solution of the
cosmological constant problem. For example, while the number of
monopoles is diluted due to inflation, the vacuum energy density
is constant during inflation.\footnote{We shall later note that
the equation of state for the vacuum is $ p = - \rho$, where $p$
is a constant due to Lorentz invariance. This equation of state,
which from a physics point of view is highly peculiar, thus gives
rise to {\em negative} pressures $ p = - \rho < 0$ corresponding
to positive energy densities $\rho > 0$. Pressures in the
Friedmann-Robertson-Walker model act contra intuitively (positive
pressures lead to deceleration of the scale factor) and thus the
occurrence of a negative pressure is needed for the outwardly
inflating universe generated by the vacuum. The vacuum equation of
state also implies that the negative pressure produces an amount
of work ($- p dV = \rho dV$) in the expanding universe, so despite
the expansion of space the energy density of the vacuum does not
decrease but remains constant.} Note that the use of vacuum energy
density in the inflation scenario is rather delicate: While
physicists would be most happy to discover some mechanism to
guarantee the vanishing of the cosmological constant {\em now},
this mechanism have better not to enforce that the cosmological
constant or vacuum energy density is always zero.

\sss{Critique of the standard QFT vacuum concept}

After having described the origin of vacuum energy in the standard
QFT picture, we shall now -- in a more critical mode -- discuss
its necessity and the question of its measurability, primarily
with focus on the QED vacuum.\footnote{We have already in previous
subsections submitted a few critical remarks concerning the QCD
vacuum and the Higgs vacuum. Both vacua are intimately associated
with spontaneous breakdown of symmetries (breakdown of chiral
symmetry and electroweak symmetry, respectively).}

Both the zero-point energy and the higher order contributions to
the vacuum energy in QED are consequences of the field
quantization. The lowest order zero-point energy of the
electromagnetic field is often removed from the theory by
reordering the operators in the Hamiltonian through a specific
operational procedure called normal (or `Wick') ordering
\cite{wick50}.\footnote{This move is mathematically justified
since the ordering of operators in a quantum theory is quite
arbitrary and is not fixed in the transition from the classical to
the quantum mechanical description of, say, the electromagnetic
field. In fact, there is also zero-point energy associated with
the free electron-positron field but this too can be removed by a
reordering of operators.} Normal ordering amounts to a formal
subtraction of the zero-point energy from the Hamiltonian, and
although this procedure does not remove the higher order
contributions to the QED vacuum energy, these can also be removed
by subtractions order by order in the perturbative expansion
(corresponding to further redefinitions of the zero for the energy
scale). 

In spite of the possibility of removing the vacuum energy, we
noted above that the experimentally demonstrated Casimir effect
seems to rest on the notion of zero-point energy. If so, the
gravitational effects of even the lowest order contribution to the
vacuum energy cannot easily be excluded in the discussion of the
nature of the vacuum. Moreover, there are other well-known
physical effects which appear to rest, not on lowest order
zero-point energy but on the reality of higher order (interacting)
vacuum fluctuations. The Lamb shift (a split of energy levels in
the Hydrogen atom), the anomalous magnetic moment of the electron
(an anomaly in the magnetic properties of the electron), and other
'higher order' effects in QED all find their theoretical
explanations through the concept of vacuum
fluctuations.\footnote{In the Feynman diagram description of
such higher order effects, the vacuum fluctuations, e.g. virtual
electron-positron pairs, are connected to external lines which
represent real particles (in contrast to the above mentioned
higher order corrections to the vacuum energy -- vacuum blob
diagrams -- which are without external lines).} Nevertheless,
these effects might be explained from a different point of view.
For instance, the Casimir effect could be ascribed to fluctuations
in the constituents of the Casimir plates rather than to a
fluctuating vacuum existing prior to the introduction of these
plates. An example of such a viewpoint is found in Schwinger's
source theory where the Casimir effect is derived without
reference to quantum fields and zero-point energy, and, according
to Schwinger, the source theory approach also avoids the notion of
vacuum fluctuations in the calculations of the Lamb shift and
other higher order QED effects \cite{rugh98}.\footnote{In a paper
by Saunders \cite{saunders99}, which is in somewhat the same
spirit as our earlier work \cite{rugh96,rugh98}, it is indicated
that Schwinger might actually be using the concept of zero-point
energy in his derivation of the Casimir effect. If correct, this
observation by Saunders -- in conjunction with the observation
(\cite{rugh98}, p. 133) that Schwinger's higher order calculations
do involve infinities, and that Schwinger (in higher orders) lets
the fields themselves be sources of fields -- of course casts some
doubt on the extent to which Schwinger really works with an empty
vacuum. Another derivation of the Casimir effect by Milonni
\cite{milonni94} also avoids zero-point energy but Milonni argues
that the QFT concept of a non-trivial vacuum is needed in any case
for consistency reasons. This argument, however, rests on the
validity of the so-called fluctuation-dissipation theorem which is
derived in a semi-classical approximation, so it is not obvious
whether the conclusion is an artefact of this approximation (see
also \cite{saunders99}).}

\ \\
{\em Further considerations on the measurability of the vacuum energy}

The question of measurability of the quantized electromagnetic fields
was addressed already by Bohr and Rosenfeld in 1933 who investigated
how accurately one can measure the components of the quantized
electromagnetic field \cite{bohr33}. They argued that the QED
formalism reflects an unphysical idealization in which field
quantities are taken to be defined at definite spacetime
points.\footnote{This unphysical idealization is, according to Bohr
and Rosenfeld, reflected mathematically in the formalism by the
appearance of infinite delta functions in the field commutation
relations.}  For measurements of the field strengths, an unambiguous
meaning can, according to Bohr and Rosenfeld, be attached only to
average values of field components over a finite spacetime region.
In the simplest realization of such measurements, an arrangement of
charged test bodies is envisaged in which the test charges are
homogeneously distributed over the finite spacetime region.  Bohr and
Rosenfeld derive an expression for the measurement uncertainty of the
field strengths showing that these diverge when the volume of the
spacetime region approaches zero (thus, field fluctuations are
ill-defined if field quantities are taken to be defined at definite
spacetime points).  A result of the investigation by Bohr and
Rosenfeld is that the origin of field fluctuations (or field energy)
in a measurement arrangement is unclear, since the result of a
measurement of the field fluctuations (or the field energy) rests on
the charged test bodies in the finite spacetime region. This is also
hinted in a letter from Bohr to Pauli commenting on the Bohr-Rosenfeld
analysis:

\begin{quote}
The idea that the field concept has to be used with
great care lies also close at hand when we remember that all field
effects in the last resort can only be observed through their effects
on matter. Thus, as Rosenfeld and I showed, it is quite impossible
to decide whether the field fluctuations are already present in 
empty space or only created by the test 
bodies.\footnote{Bohr Scientific Correspondence (film 24, section 2):
Letter from Bohr to Pauli, February 15, 1934. Translation, slightly 
modified, is taken from (\cite{kalckar96} p.34). Quoted with permission
from the Niels Bohr Archive, Copenhagen.}
\end{quote}

We have already mentioned that, since Bohr and Rosenfeld's work,
several experimentally verified physical effects -- such as the
Casimir effect, the Lamb shift, spontaneous emission from atoms
and the anomalous magnetic moment of the electron -- have been
taken to point to the reality of QED vacuum energy and (vacuum)
field fluctuations. The important point is, however, that also as
concerns these effects, the measurements are on material systems
-- the plates in the Casimir effect, the atom in the case of the
Lamb shift, etc. -- so that it seems impossible to decide whether
the experimental results point to features of the vacuum `in
itself' or of the material systems.\footnote{At earlier stages of
this investigation of properties of the vacuum, we have presented
this viewpoint as a `no-go theorem' for measurements on the vacuum
\cite{rugh96}.} Briefly summarizing the lack of experimental
possibilities of probing the vacuum in QED:

\begin{itemize}

\item 
The results of scattering experiments are explained through
calculations in which the lowest order zero-point energy is
removed by normal ordering and the higher order corrections
(vacuum blob diagrams, without external lines) are ignored as they
do not contribute to the scattering amplitudes. The vacuum energy
is therefore not associated with any physical consequences in
scattering experiments.

\item In the so-called vacuum effects, such as the Casimir effect
or the Lamb shift, it seems impossible to decide whether the
effects result from the vacuum `in itself' (the ground state of
the QED fields) or are generated by the introduction of the
measurement arrangement (or system which is measured
upon).\footnote{For example, if the Lamb shift is to be considered
as a probe of the vacuum, the atom (with its radiation field) is a
part of the `measurement arrangement' (connected, in turn, to an
external measurement arrangement which measures the emitted
frequencies of the radiation produced by this atomic arrangement).
As we have suggested previously, one might attempt to argue that a
similar situation holds as concerns the experimental pointers
toward the QCD vacuum. For example, the non-vanishing quark
condensate as an explanation for the small pion mass does not
necessarily point to features of empty space, but perhaps to
features of the material systems (the pion) studied.}

\end{itemize} 

\ \\
{\em Could observations of $\Lambda$ be an indication that there
is no real vacuum energy?}

Given that the `vacuum' effects discussed above might be explained
by referring to the material constituents of the measurement
arrangements, it could be that the most direct `experiment' on the
vacuum (i.e. the most direct probe of consequences of the vacuum)
is the macroscopic observation of the smallness or vanishing of
the cosmological constant \cite{rugh96}.\footnote{This
possibility, like the cosmological constant problem itself, rests
on the assumption that $\Lambda$ is indeed a measure of vacuum
energy. We will examine this assumption in more detail in the
following section} At first sight, it might appear that since the
measurement/observation of $\Lambda$ also involve material systems
(measurements of movements in the solar system or the dynamics of
large scale cosmological structure, etc.), it is just as indirect
a probe on the vacuum as the experimental effects discussed above.
However, there is a striking difference between the so-called
vacuum effects and the possible observed effects of a
non-vanishing $\Lambda$ on, say, the solar system. The `vacuum'
effects discussed above are all quantum effects which result from
quantum mechanical (quantum field theoretic) considerations. For
example, the Casimir effect can be calculated by considering the
quantum fluctuations of the constituents of the plates, and the
Lamb shift is calculated by studying higher order (quantum)
corrections to the radiation field between the proton and the
electrons in the atom. Such quantum effects are associated with
expressions (like $\sum \frac{1}{2} \hbar \omega$ ) in which
$\hbar$ enters. By contrast, in solar system observations, say,
the measured effect of a non-zero $\Lambda$ is revealed in the
dynamics of the test bodies in a classical gravitational field.
The Newtonian potential around the sun is for non-vanishing
$\Lambda$ modified to \cite{gibbons77,axenides00} 
\beq 
\Phi=\frac{GM}{r}+ \frac{1}{6}
\Lambda c^2 r^2 
\eeq 
where $M$ is the mass of the sun and $r$ the
distance from the sun. Such a classical observational arrangement
involving classical gravitational fields (without $\hbar$) cannot
possibly build up a quantum expression (like the expression $\sum
\frac{1}{2}\hbar \omega$) in which $\hbar$ enters.\footnote{Note
that it is not important for the measurement of $\Lambda$ in this
planetary context, that the sun and the planets are build up of
quantum constituents (whereas this is the essential feature in the
Lamb shift, the Casimir plates, etc.): If we replaced the sun and
the planet by two point masses $M$ and $m$ then this observational
arrangement could just as well detect the non-vanishing of
$\Lambda$ via the movements of those two point masses.} Thus, if
$\Lambda$ is a measure of quantum vacuum energy, one cannot assume
that the observed value of $\Lambda$ is induced by the
classical measurement arrangement.

Conventionally, it is expected that the observation of $\Lambda
\approx 0$ should be accounted for by some sort of cancellation
mechanism between the individual contributions to the QFT vacuum
energy (see the following sections). This expectation is based on
the assumption that the contributions to the QFT vacuum energy are
physical real in the sense that they have certain physical
consequences, such as gravitational effects or the attraction
between Casimir plates. But insofar as there are no clear
experimental demonstrations of the reality of vacuum energy in the
laboratory, one could also speculate that there is no real vacuum
energy associated with any of the fields in QFT. The speculation
(advanced previously in \cite{rugh96,rugh98}) that the observation
of $\Lambda \approx 0$ could indicate that there is no real QFT
vacuum energy does not necessarily imply that the standard QFT
formalism is altogether misleading. In fact, there seems to be at
least two possible interpretations of the assumption that there is
no real vacuum energy: (1) The standard QFT formalism is
maintained but one should not associate energy to fields in empty
space. The vacuum energy is therefore to be viewed as an artefact
of the theory with no independent physical existence. This view is
consistent with the fact that vacuum energy can be a practical
concept in connection with deriving quantum features, such as the
Casimir effect, of material systems -- insofar as these features
can also be accounted for by referring to the material
constituents of the systems studied. (2) The standard QFT
formalism is abandoned altogether, for instance by replacing it by
something like Schwinger's source theory. According to source
theory, there are no quantum operator fields (the fields are
c-numbered) and there are fields only when there are sources (e.g.
the material constituents of the Casimir plates). This means in
particular, as emphasized by Schwinger, that there is no energy in
empty space: ``...the vacuum is not only the state of minimum
energy, it is the state of {\em zero} energy, zero momentum, zero
angular momentum, zero charge, zero whatever." \cite{schwinger73}.

Of course, it requires further studies to determine whether (1),
(2) or something else, is a viable option for maintaining that
there is no real vacuum energy (for instance, Schwinger's source
theory only concerns QED, see also \cite{rugh98}). We suggest
therefore that it should be further examined if the substantial
conception of a QFT vacuum with non-zero energy in `empty space'
(that is, in the absence of any material constituents) involve
{\em unjustified extrapolations} beyond what is experimentally
seen.\footnote{More detailed investigations of the experimental
(and theoretical) pointers toward a substantial conception of the
QFT vacuum -- including further scrutiny of Schwinger's
alternative source theory concept -- are planned to be presented
elsewhere.} 

In the brief discussion above we have almost exclusively
considered the quantum theory of electromagnetism and the
experimental pointers toward its vacuum state. But, as we have
already seen, an examination of the experimental evidences for the
vacuum concept in quantum field theory in its wealth of
manifestations is a project which is rather involved. What does it
take to back up the general idea of `emptying the QFT vacuum for
energy' by a more detailed model building? Clearly, it is of
interest to stimulate further investigations of the concept of
spontaneous symmetry breakdown.\footnote{For instance, are the
ideas of spontaneous symmetry breakdown related to a
translational- and Lorentz invariant vacuum state of {\em
infinite} spatial extension? An apparent `spontaneous' breakdown
of symmetry also can be realized for systems with finite spatial
extension, such as in a ferromagnet.} It would also be interesting
with a discussion of alternatives to the Higgs mechanism as
generating masses of the particle content of the Standard
Model.\footnote{As we have already noted such a viewpoint has been
advanced by some physicists, e.g. Veltman and Glashow. The Higgs
mechanism, and the non-vanishing Higgs expectation value in the
vacuum, is at present implemented in the Standard Model of high
energy physics, but it does not have much explanatory power as
concerns e.g. the numerical values of the masses of the particle
constituents. The masses are essentially just parametrized by a
large number of input parameters (Yukawa couplings) adjusted to
fit experiments.} Likewise, in QCD, the viewpoint should be
examined if the pion mass, say, by necessity has to be generated
by a vacuum quark condensate existing in empty space prior to the
introduction of the physical pion systems, etc. Moreover, a
project of `emptying the QFT vacuum for energy' has potential
important bearings on how the vacuum energy is viewed in the
context of contemporary cosmological ideas in the description of
phase transitions and the idea of an early inflationary
phase.\footnote{Inflation is usually conceived to rest upon a
substantial conception of the quantum vacuum but this is not
necessarily the case \cite{brandenberger99,starobinsky80}.}

While this may seem like a high price to pay for solving the
cosmological constant problem we think, at least, that the
assumption that there is no real QFT vacuum energy should be
examined along with other explanations which have been put forward
to solve the largest discrepancy between observation and theory in
contemporary physics.

\begin{center}
\ss{General relativity and the quantum vacuum}
\end{center}

Experimentally, it is difficult to probe if there is a link
between QFT and general relativity (GR). This difficulty is
related to the smallness of the gravitational force in the micro
physical domain. In fact, whereas the link between quantum
mechanics and special relativity has lead to the experimentally
successful QFT's, we know of only very few experiments which test
the relation between any gravitational effect and non-relativistic
quantum mechanics.\footnote{A famous experiment which directly
probes the link between (Newtonian) gravity and non-relativistic
quantum mechanics is the observed gravitationally induced phases
in neutron interferometry in the experiment of Colella, Overhauser
and Werner \cite{colella75}. The experiment established the
equality of inertial mass with gravitational mass for neutrons to
an accuracy of roughly $1 \%$. Recent atomic-interferometry
experiments and neutron-interferometry experiments refine these
experiments, see also \cite{adunas00}.} But we know of absolutely
no experiments which test directly the relation between general
relativity and quantum field theory.\footnote{The Unruh-Davies
effect and Hawking radiation (see below) remain interesting theoretical
conjectures. There are proposals by Bell and Leinaas that the
Unruh-Davies effect accounts for some observations connected to
circular accelerations at CERN \cite{bell87} but alternative
explanations have been provided and it is also refuted by Unruh
(S. Habib and B. Unruh, personal communication).} However, the
cosmological constant may be such a test of the relation between a
specific quantum effect (vacuum energy) and GR.

In the previous section, we reviewed the various sources of vacuum
energy. In themselves these do not constitute a problem since any
resulting vacuum energy in QFT may be circumvented by redefining the
energy scale -- only {\em differences} in vacuum energy for various
configurations have experimental consequences.\footnote{For example,
the Casimir effect derived by means of zero-point energy is obtained
by considering the difference in vacuum energy density between two
configurations of the metallic plates, see e.g.
\cite{rugh98}.} By contrast, GR is sensitive
to an {\em absolute} value of vacuum energy. Thus, the
gravitational effect of a vacuum energy resulting from zero-point
energies, virtual particles (higher order vacuum fluctuations),
QCD condensates, fields of spontaneously broken theories, and
possible other, at present, unknown fields, might curve spacetime
beyond recognition. It is usually assumed that the vacuum energy
density ($<\rho>$) is equivalent to a contribution to the `effective'
cosmological constant in Einstein equations (\ref{Einstein1917}): 
\begin{equation} \label{Lambdaeff}
\Lambda_{eff} = \Lambda = \Lambda_{0} + \frac{8 \pi G}{c^4} < \rho_{vac} >
\end{equation}
where $\Lambda_0$ denotes Einstein's own `bare' cosmological constant
which in itself leads to a curvature of empty space, i.e. when there is no
matter or radiation present.
Once eqn (\ref{Lambdaeff}) is  established, it follows that
anything which contributes to the QFT vacuum energy density is also a
contribution to the effective cosmological constant in GR. From
our discussion above we have that the total vacuum energy density has
at least the following three contributions,
\begin{equation} \label{Energyeff}
\left(
\begin{array}{c}
\mbox{Vacuum} \\
\mbox{energy}\\
\mbox{density} 
\end{array}
\right) \; = \;
\left(
\begin{array}{c}
\mbox{Vacuum zero-} \\
\mbox{point energy}\\
\mbox{+ fluctuations} 
\end{array}
\right) \; + \;
\left(
\begin{array}{c}
\mbox{QCD gluon} \\
\mbox{and quark} \\
\mbox{condensates}
\end{array}
\right) \; + \;
\left(
\begin{array}{c}
\mbox{The} \\
\mbox{Higgs} \\
\mbox{field}
\end{array}
\right)
\; + \; \cdots
\end{equation}
\noindent
where the dots represent contributions from possible existing
sources outside the Standard Model (for instance, GUT's, string 
theories, and every other unknown contributor to
the vacuum energy density). 

There is no structure within the Standard Model which suggests any
relations between the terms in eqn (\ref{Energyeff}), and it is
therefore customary to assume that the total vacuum energy density
$\rho_{vac}$ is, at least, as large as any of the individual
terms. In order to reconcile the vacuum energy density estimate
within the Standard Model with the observational limits on the
cosmological constant (eqn \ref{obsbound}), one thus has to
`fine-tune' that $\Lambda _0$ cancels e.g. the QED contribution to
a precision of around 55 orders of magnitude if one assumes the
Standard Model to be valid up to energy scales of 100 GeV.

Before discussing some possible avenues for solutions to the cosmological
constant problem we shall address how a microscopic 
quantum energy calculated in a fixed non-curved spacetime can 
contribute to a classical equation in which spacetime is dynamical.
In discussions dealing with the cosmological constant problem, it
is sometimes merely stated that symmetry requirements implies that the
QFT energy-momentum tensor in vacuum must take the form of
a constant times the metric tensor:
\begin{equation} \label{tmn}
<0|\hat{T}_{\mu \nu}|0>= T^{vac}_{\mu\nu} = \mbox{constant} 
\times g_{\mu \nu} = 
< \rho_{vac} > g_{\mu \nu},
\end{equation} 
where the constant is identified with $ <\rho_{Vac}>$ because it
must have dimensions of an energy density. If this identification
is accepted, it follows that the huge QFT vacuum energy will act
as a contribution to Einstein's cosmological constant in
eq.(\ref{Einstein1917}), and one is thus lead to a cosmological
constant problem. However, just how symmetry constraints in
general relativity (e.g. constraints on $T_{\mu\nu}$) are related
to those imposed on a quantum vacuum state ($|0>$) is often not
discussed, so we shall attempt to clarify these points which
involve the difficult subject of quantum field theory in curved
spacetime backgrounds. 

\ \\
\sss{The vacuum energy-momentum tensor in general relativity}

Disregarding for the moment quantum properties of the
vacuum, we seek the energy characteristics of the vacuum in the 
form of an equation for a relativistic covariant energy-momentum tensor
$T_{\mu\nu}^{vac}$. 
As we will see below, we are interested in the case where
the (quantum) vacuum state is Lorentz invariant 
in Minkowski spacetime. In special relativity,
a Lorentz invariant $T^{vac}_{\mu\nu}$ implies that it must have the form
\beq
\label{minkowski}
T_{\mu \nu}^{vac} = \mbox{constant} \times \eta_{\mu \nu}
\eeq
where $\eta_{\mu \nu}$ is the Minkowski 
metric.\footnote{If it is assumed that the vacuum is characterized by a
non-zero energy density $\rho_{vac}$, one might
ask how is it possible that the vacuum state is strictly
relativistic invariant? At first, one would not be able to
construct a physical state with a definite non-vanishing
energy-momentum 4-vector $ \{ E, \vec{p} \} $ which is
Lorentz invariant. For example, $ \{ E, \vec{0} \} $
would yield $\vec{p} \neq 0$ in another reference frame (unless
the energy $E$ has been put to zero as well). However, whereas it
is not possible to construct a Lorentz invariant state with a
non-vanishing energy-momentum vector, it is indeed possible to
construct a physical state with a non-vanishing Lorentz invariant
energy-momentum {\em tensor} (that is specifying a {\em density}
of energy and momentum, rather than absolute values).}
It is instructive
to compare this $T_{\mu\nu}$ to that of  
a relativistic perfect fluid
\beq
\label{perffluid}
T_{\mu\nu}=(p+\rho)u_{\mu}u_{\nu}+p \; \eta_{\mu\nu}
\eeq
from where we see that the vacuum can be mathematically
characterized as a `perfect fluid' with the equation of state 
$p=-\rho$.
After establishing the equation (\ref{minkowski})
in flat spacetime, it is first expressed in
an arbitrary coordinate frame (looking at equation
(\ref{minkowski}) in flat
spacetime from general coordinates) so that $\eta_{\mu \nu}$ is
replaced with $g_{\mu \nu}$. Then, using the 
principle of general covariance
(see e.g. \cite{weinberg72} p. 92)
it is asserted that the form of $T_{\mu \nu}^{vac}$ has to
be a $\mbox{constant} \times g_{\mu \nu}$ also in the general
case where 
$g_{\mu \nu}$ describes a real gravitational field (with non-vanishing
Riemann tensor field components, e.g. for a curved model of the
universe). Following Einstein, this is a standard algorithm for 
incorporating the effects of gravitation on
physical systems, cf. (\cite{weinberg72} p. 105-106): Write the
appropriate special-relativistic equations that hold in the
absence of gravitation, replace $\eta_{\mu \nu}$ with
$g_{\mu \nu}$, and then replace all derivatives with
covariant derivatives.

\sss{The quantum vacuum in various spacetime backgrounds}

We shall now address how much of the above discussion can be taken
over when one attempts to calculate $T^{vac}_{\mu\nu}$ as a vacuum
expectation value of the quantum operator $\hat{T}_{\mu \nu}$ in
various spacetime backgrounds. In particular, we are interested in
whether or not we in all circumstances can justify equation
(\ref{tmn}), for instance whether the quantity $<0|\hat{T}_{\mu\nu}|0>$ is
well-defined. As mentioned above, such a justification is crucial
for formulating the cosmological constant problem.

\ \\
{\em Vacuum in Minkowski spacetime}

As long as we are in Minkowski spacetime, Lorentz invariance of
the vacuum state is build into the QFT formalism.\footnote{Besides
Lorentz invariance, it will in flat Minkowski spacetime be
`natural' to assume the additional constraint of invariance on the
vacuum state under translations in time and space. Thus, in
Minkowski spacetime the vacuum state is invariant under the entire
Poincare group (also called the inhomogeneous Lorentz group). As
we have seen already, however, the vacuum energy is envisaged to
change during the phase transitions in the universe, so the vacuum
is clearly not invariant under time translations (which is also
evident since the Big Bang model of the universe is described by a
Robertson-Walker metric rather than a Minkowski metric). Poincare
invariance is therefore not a symmetry fulfilled by the vacuum
state (of the various quantum fields) in the actual universe.
Moreover, the ground state of finite temperature QFT is not even
Lorentz invariant due to the presence of excitation quanta in this
state. However, as we indicated above, one usually distinguishes
the thermal energy of the quanta from the energy of the `vacuum'
part of this ground state. The latter part of this state is
Lorentz invariant and may, as Guth argued, drive inflation.} Thus
if $|0>$ is a vacuum state in a reference system $\cal{R}$ and
$|0'>$ refers to the same vacuum state observed from a reference
frame $\cal{R'}$ which moves with uniform velocity relative to
$R$, then the quantum expression for Lorentz invariance of the
vacuum state reads
$$|0'>\; =\; {\cal{U}}(L)|0>\;=\;|0>$$
where ${\cal{U}}(L)$ is the unitary transformation (acting on
the quantum state $|0>$) corresponding to a Lorentz transformation $L$
(see e.g. \cite{weinberg96}, chapt. 2).
All physical properties extracted from this vacuum state, such as the 
expectation value of the energy momentum tensor,
should also remain invariant under Lorentz transformations 
(i.e. $<0'|T_{\mu\nu}'|0'> = <0|T_{\mu\nu}|0>$). This symmetry
requirement can only be fulfilled if
$$<0|\hat{T}_{\mu \nu}|0>= \mbox{constant} \times \eta_{\mu\nu}$$
so that eq.(\ref{tmn}) is indeed satisfied.  

These aspects of Lorentz invariance of the vacuum state also serve to
clear up another conceptual issue, namely the question of why, given
all the vacuum fluctuations, particles 
cannot be scattered on this vacuum state. For example, it is not 
possible to scatter a photon (an electromagnetic
wave) on vacuum fluctuations. This can be shown by using 4-momentum
considerations and the condition of Lorentz invariance of the
vacuum state (if this was not the case, an `empty space' comprised
of vacuum fluctuations etc. would make optical astronomy
impossible).\footnote{As we have noted earlier, one {\em can} scatter on
`ordinary' quantum mechanical zero-point fluctuations in a material 
system (the ground state of which is not Lorentz invariant).}

\ \\
{\em  Vacuum in flat spacetime}

When the spacetime is still flat but not Minkowski, the Minkowski
metric is transformed into a non-trivial spacetime metric
$g_{\mu\nu}$. In some cases of non-trivial spacetimes it is
difficult to identify a vacuum state as the state of no particles,
hence it is not clear that a vacuum state $|0>$ and therefore
$<0|\hat{T}_{\mu\nu}|0>$ are well-defined. A most striking example
of this situation is provided by the Unruh-Davies effect which
predicts that an accelerated particle detector moving through the
Minkowski vacuum will detect particles (Birrell and Davies
\cite{birrell82} p.54). Nevertheless, it can still be argued that
it is the acceleration which creates the particles so that the
natural vacuum state, after all, can be taken as the Minkowski
vacuum as experienced by inertial (non-accelerated) observers. On
this view, the experiences of an accelerated observer is
attributed to the experiences of these observers ``being
`distorted' by the effects of their non-uniform motion''
(\cite{birrell82} p.55). Eqn. (\ref{tmn}) can therefore be upheld
if we select a non-accelerated reference frame. 

\ \\
{\em Vacuum in curved spacetime}

The gravitational field (e.g. in our expanding universe) will in
general be expected to produce particles, thereby obscuring the
concept of vacuum as a state with no particles. In fact, when
gravitational fields are taken into account ``...inertial
observers become free-falling observers, and in general no two
free-falling detectors will agree on a choice of vacuum''
(\cite{birrell82} p.55). So whereas the state $|0>$ with no
particles is the obvious vacuum state in Minkowski spacetime, in
general there is not any reference system in which there is no
particle production. With no clear vacuum concept, one cannot give
a precise meaning to $<0|\hat{T}_{\mu\nu}|0>$, let alone associate
this quantity with a cosmological term as in eq.(\ref{tmn}). 
These issues can be further examined by writing down the central
equation for discussions of QFT's in curved spacetime backgrounds
(e.g. \cite{birrell82} p.154):
\beq
\label{curvedGR}
R_{\mu \nu} - \frac{1}{2} g_{\mu \nu} R -
\Lambda g_{\mu \nu} =  \frac{8 \pi G}{c^4} \; <\hat{T}_{\mu \nu}>
\eeq
where the notation is the same as in eq.(\ref{Einstein1917}), but
where the right hand side is now the expectation value of a
quantum operator for relevant states (e.g. the vacuum state). This
semi-classical equation assumes a specific relation between the
{\em classical} gravitation field and the {\em quantum}
expectation value of $<T_{\mu\nu}>$. Conjectured applications of
eq.(\ref{curvedGR}) involve both more local physical phenomena,
e.g. involving quantum fields in the vicinity of black holes
(Hawking radiation) as well as global properties of quantum fields
in the entire universe (e.g. particle production due to spacetime
curvature in an expanding universe). As we noted above, however,
there has so far been no experiments or observations which test
the GR-QFT relationship and, specifically, eq. (\ref{curvedGR}) has
not been tested.\footnote{There are, of course, various motivations
for setting up eq.(\ref{curvedGR}) based e.g. on the analogy with
``...the successful semi-classical theory of electrodynamics, where
the classical electromagnetic field is coupled to the {\em expectation
value} of the electric current operator" (\cite{birrell82} p.154).} 

Apart from the issue of experimental support of
eq.(\ref{curvedGR}), the question is whether the equation is
meaningful: First, it is difficult to get a sensible value for
$<\hat{T}_{\mu \nu}>$ since there are more divergences introduced
in the curved spacetime case relative to the flat spacetime case,
so that $<\hat{T}_{\mu \nu}>$ remains divergent even if
$<\hat{T}_{\mu \nu}>_{flat}$ is subtracted (\cite{birrell82}
p.153). Furthermore it is only possible for some simple cases
(e.g. quantum fields in a static universe) to calculate
renormalized (finite) values for $<\hat{T}_{\mu
\nu}>$.\footnote{The question of regularizing QFT's in curved
spacetime has been discussed further since Birrell and Davies
\cite{birrell82}. For instance, developments in the so-called zeta
function regularization programme, promoted e.g. by Hawking
already in the late 1970s, have suggested this method to be one of
the more fruitful ways of removing divergences for QFT's in curved
spacetime (see e.g. \cite{elizalde95}). The situation, however, is
still not clear -- for instance one would not like physical
results to depend on particular regularization schemes. We thank
F. Antonsen for discussions on this subject.} Second, and more
general, eq.(\ref{curvedGR}) represents a `back-reaction' problem
which needs to be solved in some self-consistent way (the
divergence difficulties just mentioned still assume a fixed curved
background): If $<\hat{T}_{\mu \nu}>$ affects the metric, then
this metric will change the assumptions for calculating
$<\hat{T}_{\mu \nu}>$ (see e.g. \cite{wald94}
p.54).\footnote{Further discussion of the problems with eq.
(\ref{curvedGR}) can be found e.g. in (\cite{wald94} p.98.)} This
`hen and egg' problem is complicated not just because
back-reaction problems in general are difficult to
solve, but because the 
interesting physical states, e.g. the vacuum state $|0>$, 
appearing on the right hand side of the equation are hardly 
well-defined in curved spacetime except in certain special cases!

Finally, when interactions are taken into account in the curved
space case, it becomes even more difficult to establish general
renormalizability of $<\hat{T}_{\mu \nu}>$ (\cite{birrell82}
p.292), which means that we do not know if, for example,
interacting QED is renormalizable (and thus physical meaningful)
in curved space:

\begin{quote}
Will a field theory (e.g. QED) that is renormalizable in 
Minkowski space remain so when the spacetime has a non-trivial
topology or curvature? This question is of vital importance,
for if a field theory is to lose its predictive power as soon
as a small gravitational perturbation occurs, then its physical
utility is suspect. It turns out to be remarkably difficult
to establish general renormalizability... (Birrell and Davies 
\cite{birrell82} p.292).
\end{quote}

However, even if general renormalizability is difficult to
establish, it seems that since the empirical support of QED has
been obtained in laboratory frameworks where a small gravitational
field has indeed been present, the loss of predictive power cannot
be fatal. Indeed, the stability of QED predictions (the Lamb
shift, the anomalous magnetic moment of the electron, etc.) under
small gravitational perturbations, e.g. due to the earth's gravitational 
field, have been quite remarkable.

\ \\
{\em Is there still a cosmological constant problem?}

From the above discussion, it might seem that it is no longer clear
that a cosmological constant problem can be formulated
when quantum fields are considered in a curved spacetime on
which they back react. As we have seen, the formulation
is problematic both at a technical and at a conceptual
level. Technically, it is extremely difficult to calculate
$T_{\mu \nu}$ in a given fixed background and incorporation
of the back reaction effects makes it close to impossible. 
Conceptually, the very notion of a vacuum is not well-defined
in a curved spacetime background. 

However, the technical and conceptual shortcomings in formulating
the cosmological constant problem in a precise mathematical sense
are not sufficient to exclude that modern physics is faced with
an intriguing problem. From observations in astrophysics it is known
that the gravitational field in our local neighbourhood (our solar
system etc.) is rather weak, i.e. in suitably chosen coordinates
the spacetime around us may be written as the Minkowski spacetime
metric plus a small perturbation from weak gravitational fields.
As noted above, the stability of the predictions of QFT under the
influence of such small gravitational perturbations thus makes it
reasonable to expect that we can apply, with some approximate
accuracy, standard QFT formulated in Minkowski spacetime in our
local astrophysical neighbourhood. We are then, at least in this
local domain of the universe, faced with a QFT prediction of an
enormous energy density which nevertheless has no visible
astrophysical (gravitational) effects.

Moreover, astrophysical evidence is consistent with an expanding
Robertson-Walker metric on cosmological scales where the expansion
is rather slow and the metric close to spatially flat. Under these
circumstances, it may be possible to define an approximate vacuum
state of the quantum fields in the universe which has almost no
particles in it (\cite{birrell82} p.63-65, 70). From Birrell and
Davies discussion it is clear that the question of how well a
vacuum state can be defined in curved spacetime is still not
settled in detail (different authors get different results) but
they seem themselves to get a reasonable definition of the vacuum
state in which there, at present, is less particle creation than
16 particles per km$^3$ per year (\cite{birrell82} p.73)!

Whatever the merits of these (difficult) attempts of defining an
exact vacuum state in an expanding universe, it seems in any case
as if the observational input of a quasi-flat, slowly expanding,
universe to some extent justifies that one ignores the problems of
QFT in curved spacetime. In this sense {\em observations} help to make
the cosmological constant problem reasonably well-defined.\\

\begin{center}
\ss{Review and analysis of possible solutions}
\end{center}

Since the revival of the cosmological constant problem, at least
since the early 1980s, it has been regarded as a fundamental
problem in modern physics.\footnote{In the discussion section
which follows we shall come back to why there nevertheless have
been, and still are, differences in physicists conception of the
problem.} With this state of affairs it is interesting to outline
where the problem could reside. For this reason we suggest a
conceptual (somewhat schematic) classification of possible
solutions types. Most of these solution types are discussed in
much more detail elsewhere (e.g. \cite{Weinberg89,sahni99}) but
we indicate below the ideas on which they are based. We have
already emphasized that the cosmological constant problem is not a
problem for QFT or GR in isolation but emerges when these two
theories are considered together. Nevertheless, the origin of the
problem could reside in either of these theories, as well as in
the link between them. Logically then, it appears that there are
three possible solution types to the problem (although there is
some overlap between the categories, and thus that these should be
considered as heuristic guides rather than rigorous definitions):
\\

\begin{enumerate}

\item {\em A modification of GR}. The problem could be either 

\begin{enumerate}

\item `internal' in the sense that a change is needed
in the GR formalism itself (e.g. changing the role of the 
metric), or

\item `external' in the sense that GR is still considered 
effectively correct, but that it needs to be  embedded in an
extended framework to address the question (e.g. quantum cosmology).

\end{enumerate}

\item {\em A modification of QFT}. Again, the problem could be either

\begin{enumerate}

\item `internal' in the sense that
a change in, or a reinterpretation of, the QFT formalism
which gives rise to the vacuum energy is needed
(for instance through Schwinger's source theory), or

\item `external' in the sense that QFT (the Standard Model) 
is considered effectively correct as a low energy theory, but needs 
to be embedded in an extended framework to address the question 
(e.g. supersymmetry).

\end{enumerate}

\item {\em The link between GR and QFT is problematic}. Once more,
we see at least two ways in which this may be the case, either
the problem is 

\begin{enumerate}

\item `internal' in the sense that the link cannot even be
discussed properly due to our limited understanding of the coupling between 
GR and QFT (e.g. QFT in curved spacetime,
and back-reaction), or

\item `external' in the sense that we due to the limited
understanding of the coupling between GR and QFT ought to postpone
the problem until we have a theory in which the link is embedded
in an extended framework for both GR and QFT since only in such a
theory will the problem be completely well posed (e.g.
string theory).\footnote{The diagnoses of the problem for 3.(a) and
3.(b) are, however, very similar so `internal´ and `external'
should be read even more heuristically than in the first two
categories.}

\end{enumerate}
\end{enumerate}

Following this classification, we shall briefly comment on
some examples illustrating the various solution types. 
 We note, however, that no consensus exists
as to whether suggestions along any of these lines provide a solution
to the cosmological constant problem. 

It has been suggested to change the Einstein theory of gravity
itself -- solution type 1.(a) -- so that not all metrical degrees of
freedom are treated as dynamical degrees of freedom (several such
modifications of GR are possible). Weinberg notes that such
proposals do ``...not solve the cosmological constant problem, but
it does change it in a suggestive way" (\cite{Weinberg89} p.11).
Recently, it has been very popular to speculate that the
cosmological `constant' in Einstein's equation may in fact be
time-dependent (not just at phase transitions), or even represent
a new type of matter in the universe with no, or very weak, coupling
to the fields in the Standard Model. Various models of this type, often 
referred to as `quintessence' models, which may also be classified as 
type 1.(a), are discussed in \cite{sahni99} (see also 2.(b) below).

Solutions of type 1.(b) usually involve attempts to quantize gravity,
resulting in a theory of quantum gravity where classical GR comes out
in certain limits.  According to Weinberg, a decade ago, the
understanding of the cosmological constant problem in the framework of
ideas from quantum cosmology -- where quantum mechanics is applied to
the whole universe -- appeared to be `the most promising'
(\cite{Weinberg89} p. 20). Very famous are the ideas of Hawking and
Coleman about baby universes and wormholes, suggesting a probability
distribution for the cosmological constant to be peaked around
zero. However, it is widely admitted that the procedures involved are
mathematically ill-defined, including ill-defined (unbounded) path
integrals.  Whereas Coleman et al. above are making assumptions about
the structure of quantum gravity in the far ultraviolet (e.g. at
Planck scales), it has been argued by some authors that the
cosmological constant problem is an `infrared' problem, and knowledge
about quantum gravity effects extracted in the infrared, low energy,
region is sufficient to produce a small cosmological constant today
(see e.g. \cite{tsamis93,mottolaXX}).\footnote{Note that this
argument, if correct, may justify that one disregards the ultraviolet
divergence problems in eq.(\ref{curvedGR}).} For example, a
calculation of the effective damping of $\Lambda$ by
second order infrared quantum gravity effects is presented in
\cite{tsamis93}.

As concerns type 2.(a), we are not aware of any published solution
proposals along this line, apart from indications in our own
investigations \cite{rugh96,rugh98} and in a preprint in
preparation by Saunders \cite{saunders99}. Reasons for this
probably include the enormous success that QFT has had, which
mutes the motivation for searching for alternatives. However, as
we have discussed above, the cosmological constant problem might
be a motivation to search for alternative interpretations of QFT
in which the vacuum energy is not seen as physical real (and we
have indicated that experiments interpreted as showing the vacuum
energy to be real might in fact probe properties of the material
systems rather than empty space itself). Furthermore, we indicated
that Schwinger's source theory claims to work with a completely
empty vacuum (recall, however, the critical remarks on Schwinger's
theory noted earlier).\footnote{To our knowledge, neither
Schwinger nor his source theory students have discussed the
cosmological constant in a source theory context.} In any case, to
device a (dis)solution of the cosmological constant problem by
either reinterpreting the QFT vacuum energy as physical unreal or
replacing QFT by Schwinger's source theory (or another theory
working with an empty vacuum), one would have to scrutinize
further also the vacuum concept associated with QCD and the
electroweak spontaneous symmetry breaking.

Supersymmetry -- solution type 2.(b) -- embeds the standard model
of electroweak and strong interactions in an extended framework in
which each particle has a superpartner. In a supersymmetric theory
the fermion and boson contributions to the vacuum energy would
cancel to an exact zero (they are equally large and
have opposite sign), so if we lived in a world in which each
particle had a superpartner, we would understand why the vacuum
energy vanishes. However, one does not observe such
superpartners in nature, so the supersymmetry must be broken. Thus
the above mentioned cancellation no longer takes place, and the
vacuum energy density of the theory will be non-zero and
large.\footnote{Some authors speculate if we can ``...somehow
reinterpret the real world in terms of {\em unbroken}
supersymmetry, suitably constructed, even though the boson and
fermion masses are different?'', Witten \cite{witten00}.} The
construction of even more `all encompassing' supersymmetric
theories, including supersymmetric extended frameworks for the
gravitational sector (either supergravity or superstring
theories), solution type 2.(b) or 3.(b), have been attempted but
until now these theories do not seem to offer a solution to the
cosmological constant problem either (see e.g. Weinberg's
discussion \cite{Weinberg89} p. 5-6). Also along the lines of
type 2.(b), various models have been advanced in which extra 
(unobserved, weakly coupled) scalar fields
are introduced and coupled to the rest of the theory so as to make
up for `adjustment mechanisms' to cancel the cosmological constant
without obvious fine tuning being introduced. While Weinberg
remains sceptical (\cite{Weinberg89}, pp. 9 - 11), Dolgov sees
such mechanisms as the most promising candidate for a solution to
the cosmological constant problem \cite{dolgov97}. 

As for type 3.(a), the idea that the understanding of the coupling
between QFT and GR is insufficient to pose the problem, is
effectively a way to diminish the importance of the cosmological
constant problem. As we discussed above it may be questioned, for
instance considering the lack of experimental support, if the
semi-classical equation (eq.(\ref{curvedGR})) in which quantum
energy acts as a source of the gravitational field is valid.
Moreover, we have mentioned the question of whether the quantities
(e.g. the concept of a vacuum state) appearing in
eq.(\ref{curvedGR}) are well-defined. A further problem (a
`hen-egg' problem) was what comes first: The background geometry
which depends on $<T_{\mu\nu}>$, or $<T_{\mu\nu}>$ which depends
on the background geometry? On the other hand, we have pointed out
that the observational input of an approximately flat spacetime
(due to the overall low density of matter in the universe as well
as the observed smallness or vanishing of the cosmological
constant) to some extent makes it reasonable to ignore the effects
of treating the quantum fields in curved spacetime. In this sense
it appears that we need the observation of an almost vanishing
cosmological constant in order to make the cosmological constant
problem reasonably well-defined. 

For solution type 3.(b) it should be noted that the most
elaborated theory involving both GR and QFT, string theory, has so
far failed to give a plausible answer to the puzzle (but see also
type 1.(b) discussed above). According to
Witten \cite{witten00}:
\begin{quote}  
As the problem really involves quantum gravity, string theory
is the only framework for addressing it, at least with our
present state of knowledge. Moreover, in string theory, the 
question is very sharply posed, as there is no dimensionless
parameter. Assuming that the dynamics gives a unique answer
for the vacuum, there will be a unique prediction for the
cosmological constant. But that is, at best, a futuristic
way of putting things. We are not anywhere near, in practice,
to understanding how there would be a unique solution for
the dynamics. In fact, with what we presently know, it seems
almost impossible for this to be true...
\end{quote}
Thus, it is not clear that a solution to the cosmological constant
problem can be found within the framework of string theory. In
fact, the cosmological constant problem has historically been a
main obstacle of making string theory more realistic
(\cite{witten97} p.274-5) and it appears that the problem
continues to haunt this theory.\footnote{See also Iengo and Zhu
\cite{iengo99}, and (\cite{cao99} p.385,386) where Gross remarks
that, in string theory, "one cannot fudge it [the cosmological
constant problem] as in previous theories".} 

\ \\
{\em Anthropic considerations}

We finally mention anthropic considerations which fall somewhat
out of our classification scheme for solutions to the cosmological
constant problem. Since anthropic arguments seem to attract
considerable interest in this context, and since they involve
philosophically unorthodox elements, we shall discuss the these
arguments in slightly more detail. 

The proposal of an anthropic solution to the cosmological constant
problem mostly concerns the idea that our universe is embedded in
a larger structure (a `multi-verse'), and that we live in a
universe in which the cosmological constant is compatible with
conditions for life forms to evolve
\cite{Weinberg89,weinberg96b}.\footnote{Depending on which type of
multi-verse scenario is envisaged, the anthropic solution is
related to type 1.(b) (embedding GR in a broader, quantum
cosmological, context) or 2.(b) (embedding QFT in a broader,
inflationary, context) mentioned above.} If one only varies the
cosmological constant -- and thus enforce the natural laws and all
other constants of nature to remain fixed -- it is rather easy to
verify that it has to be close to zero and within the observed
upper bound to within some few orders of magnitude or
so:\footnote{Only to vary one single parameter, and keep the rest
of the structure (natural laws + other parameters) fixed, is
almost always employed (see also \cite{nielsenrugh}) in the
mathematical investigations backing up claims about the
implementation of a so-called (weak) `anthropic cosmological
principle' in our actual universe -- see below.} If the
cosmological constant is positive and too large, the universe will
too early enter an expanding phase without the formation of
sufficiently large gravitational condensation (no formation of
galaxies, stars and planets etc.). If, on the other hand,
$\Lambda$ is negative and too large (numerically) the entire
universe will re-collapse too fast, so that stars and planets do
not have time to evolve before the universe has re-collapsed.

The ``anthropic principle" covers a spectre of different versions
which, in Weinberg's words ranges from ``those that are so weak as
to be trivial to those that are so strong as to be absurd"
\cite{Weinberg89}. Nevertheless, even for a `moderate´ (or weak)
version of the anthropic principle -- stating e.g. that ``our
location in the Universe is necessarily privileged to the extent
of being compatible with our existence as observers"
\cite{carter74} -- there are quite different opinions among
physicists and cosmologists about the scientific status of such a
principle. For instance, while Weinberg has continually made use
of anthropic reasoning in connection with the cosmological
constant problem, and presently sees the anthropic line of
reasoning to the cosmological constant problem as the most
promising (see e.g. \cite{weinberg96b} and \cite{cao99} p.385),
other physicists finds this mode of thinking less
convincing.\footnote{For instance, it is noted in Kolb and Turner
(\cite{kolb93} p.269): ``It is unclear to one of the authors how a
concept as lame as the `anthropic idea' was ever elevated to the
status of a principle''.}

We shall not here attempt a detailed evaluation of the anthropic
principle in any of its forms but merely note that the resort to
anthropic considerations seems to imply a number of questionable
moves, which are primarily of a philosophical nature. Indeed,
anthropic reasoning seems to radically change what it means to
give a scientific explanation within the physical sciences. How
can the fact that the universe is hospitable to observers
constitute an explanation of anything? A detailed criticism on the
role of explanation in connection with anthropic reasoning is
given in a review of the anthropic principle by Earman
\cite{earman87}. However, as also noted by Earman, the principle
might have some explanatory power if applied in the multi-verse
scenario provided e.g. by some versions of inflationary cosmology.
This is precisely the type of scenario in which for instance
Weinberg discusses the cosmological constant
problem.\footnote{Note that anthropic considerations have not been
restricted to speculations of the cosmological constant, but have
also been invoked in an ``explanation" that other constants of
physics could not have been much different in our universe if life
is to appear in it. For a extensive list of references on the
anthropic principle, see Balashov \cite{balashov91}.} 

The use of multi-verse scenarios, however, leads to other worries
with anthropic reasoning. If a solution to the cosmological
constant problem was devised by appealing to anthropic reasoning
in a multi-verse scenario, employed e.g. by certain models of the
inflationary universe, this would undermine the usual
observational basis for a scientific explanation: One is {\em in
principle} unable to verify the existence of an ensemble of
universes observationally since we cannot, per definition, be in
causal contact with these other universes as they lie outside our
horizon (outside our light-cone). One might argue that if a
multi-verse model (e.g. a version of inflation) is supported by
means of observations in our universe, it is not unfair to
speculate on conditions in other universes. Nevertheless, Weinberg
(for instance) still has to make {\em a priori} assumptions about
probabilities of values of the vacuum energy in other universes in
order to draw a statistical conclusion that we, as typical
observers, inhabit a low $\Lambda$ universe \cite{weinberg96b}.
Due to the inobservability of these hypothetical other universes,
this effectively amounts to the curious situation of making
statistical arguments based on only one data point (the conditions
in our Universe). 

Regardless of the, at least, unclear status of anthropic reasoning
for providing a scientific explanation of the cosmological
constant, the very fact that it is mentioned as a possibility in
almost every review of the problem may be just another pointer to
how serious the cosmological constant problem is conceived to be. \\

\begin{center}
\ss{The status of the cosmological constant problem}
\end{center}

It is clear that the physicists conception of the cosmological
constant problem has been changing significantly over the years,
from Pauli's ``amusement" over coffee in the 1920s to the modern
view that the cosmological constant is a fundamental problem in
physics. As we have described, this changing historical conception
includes (1) the realization by Zel'dovich in the late 1960s that
zero-point energy cannot be ignored when gravity is taken into
account (2) The appearance of spontaneous symmetry breaking and
its possible implication for the early universe (early-mid 1970s),
(3) the advent of inflationary cosmology, based specifically on
vacuum energy (early 1980s), and (4) the realization that a
non-vanishing cosmological constant is a main obstacle for making
string theories more realistic (mid 1980s, cf. \cite{witten97}).

Within the last 10 years the cosmological constant problem has
been labelled from an ``unexplained puzzle'' (Kolb and Turner
1993, \cite{kolb93} p.198) to a ``veritable crisis'' (Weinberg
1989, \cite{Weinberg89} p.1)) to ``the most striking problem in
contemporary fundamental physics'' (Dolgov 1997, \cite{dolgov97}
p.1). While these statements differ in emphasis, which might in
part be explained by the more or less trivial psychological point
that people working on a particular problem tend to emphasize its
importance, all of the authors agree that there is a problem to be
solved -- although there is little agreement about the direction
in which a solution should be sought.

Some physicists have assumed that the connection between QFT and GR,
and in particular the cosmological constant problem, can only be
properly addressed in the context of some more general framework of a
quantum gravity theory, though there are disagreements as to whether
it is sufficient to know properties of such a quantum gravity theory
at low energies (in the infrared limit) or if it is necessary to
incorporate conjectures about the behaviour of the quantum gravity
theory at fundamental scales, such as Planck scales (the ultraviolet).
The currently most studied example of a quantum gravity theory is that
of string theory but, as we have seen, the cosmological constant
problem is regarded as a major theoretical obstacle in making progress
of such a theory.

A different view, although not very common, is that there might
not be anything in need of an explanation. This appears to be what
Bludman and Ruderman suggested with their terminology of a
``pseudo-problem" -- that one might simply take the view that the
value of the (observed) cosmological constant is a new fundamental
constant which might not be derivable from some fundamental
theory. As Bludman and Ruderman note, one can get any value for
the vacuum energy in QFT (hence also the value zero) by adding
suitable counter terms to the Lagrangian. Coleman has remarked,
however, that such an idea is not an attractive addition to
physics (\cite{cao99} p.386):
\begin{quote}
...the cosmological constant is the mass of a box of empty space,
You can always fine-tune it to zero. And nobody will say you can't
do it, but nobody will applaud you when you do it, either.
\end{quote}

\sss{The naturalness of the cosmological constant problem}

In order to further clarify the status of the cosmological
constant problem, it would at first seem instructive to compare
with occurrences of similar ``crises" and their solutions
throughout past history. At second sight it appears to us,
however, that each crisis and its solution is of rather individual
character and only with caution should one rely on analogies and
similarities drawn between the cosmological constant problem and
such previous occurrences.\footnote{Thus, we do not intend to fit
the notion of crisis in any preconceived scheme such as Kuhn's.} 

Nevertheless, Abbott attempts to make use of two historical
analogies in order to address the nature of the cosmological
constant problem \cite{Abbott88}. These analogies are (1) the
relation between the ether drift velocity and the velocity of the
earth, and (2) the mathematical relationship between the velocity
of light, the vacuum permitivity, and the vacuum susceptibility
$$c=\frac{1}{\sqrt{ \epsilon_0 \mu_o} } \; $$ which was
established numerically before the advent of Maxwell's
electromagnetic theory of light. According to Abbott, the first
relation is `unnatural' since it involves a range of unknown
parameters (e.g. the parameters describing the velocity of the
earth relative to the distant galaxies). By contrast, the second
relation involves a few well-known parameters and is therefore
`natural'. Abbott's point is that the `unnatural' relation
historically revealed itself as being based on a misunderstanding
(there is no ether -- hence no ether drift velocity), while the
`natural' relation inspired Maxwell to develop his unified theory
of radiation and electromagnetism. Since the equation $\Lambda =
0$ also involves a number of unknown quantities (e.g. the bare
cosmological constant and the terms from as-yet unknown fields,
recall eqn \ref{Energyeff}), it is best classified as an
`unnatural' relation and hence it can be expected that is covers a
misunderstanding (the complicated QFT vacuum?) rather than the
signpost of a unified theory. Nevertheless, as Abbott notes, to
clear up this misunderstanding ``without destroying the towering
edifice we have built on it [the successes of QFT]'' is a hard
challenge.

Despite Abbott's examples, it is of course far from obvious that
natural relations always point toward unified theories while
`unnatural' relations cover the fact that something is deeply
misunderstood. In any case, it it interesting that Zel'dovich 
did try to construct a numerical value for the cosmological constant
as a `natural relation' from the perspective of the dream of
unification between the macroscopic and the microscopic domain. In
much the same spirit as Eddington and Dirac's search for
coincidence between large numbers in cosmology, Zel'dovich
(\cite{zeldovich68} p.384) constructs the cosmological constant as
the combination
\beq 
\label{natural}
\Lambda \sim G^2 m_p^6/ \hbar^4
\eeq
where $m_p$ is the proton mass (assumed in those
days to be a fundamental particle), and $G$ and $\hbar$ are
Newton's gravitational constant and Planck's constant, respectively.
However, as Zel'dovich notes, this `natural relation' (\ref{natural})
is approximately 7 orders of magnitude larger than the
observational constraint. But, Zel'dovich continues (\cite{zeldovich68}, 
p. 392):
\begin{quote}
``Numerical agreement could be obtained by replacing $m_p^6$
with $m_p^4 m_e^2$, or by choosing other
powers and replacing $\hbar c$ with $e^2$, this is
essentially what Dirac and Eddington did. However, even
a discrepancy of ``only'' $10^7$ times is an accomplishment
compared with the discrepancy of the estimates by a
factor $10^{46}$.''
\end{quote}
In spite of Zel'dovich's optimism it appears that it has not, so
far, been possible to establish `natural' relations for the
cosmological constant. But it is, as already indicated, not
clear if the classification of the cosmological constant problem
as involving a `unnatural' relation is an unambiguous pointer to
its status, let alone its solution.

\sss{Some concluding remarks} 

We have attempted to clarify the origin and development of the
understanding of the cosmological constant problem as constituting a
fundamental problem in physics. At the same time we have argued that
the conception of a problem -- although motivated by observations and
theoretical expectations -- involves at least two (partly
philosophical) convictions.\\

\noindent
{\em 1. The quantum vacuum energy is physical real}

\noindent One must be convinced that the various QFT contributions
to the vacuum energy density indeed result in a physical real
energy density of empty space. While this conviction appears
natural, at least in the context of QED and QCD, due to the
apparent experimental demonstrations of the reality of various
vacuum effects, we have hinted that this conclusion could be
ambiguous. In particular, we indicated that the QED vacuum energy
concept might be an artefact of the formalism with no physical
existence independent of material systems. One possible way to
maintain such a viewpoint would be to replace QED with Schwinger's
source theory, insofar as this theory can explain QED experiments
without recourse to vacuum energy. But, regardless of the merits of
source theory, the fact that all QED (and QCD) `vacuum'
experiments involve material systems makes it reasonable to
question whether such experiments are useful to predict how empty
space `in itself' will curve spacetime. This is not meant to be a
repetition of the well-known Kantian doctrine that one cannot
obtain information of things `in themselves'. Our point is that
since the cosmological constant in Einstein's equations is a direct
measure of how much empty space `in itself' will curve spacetime,
the `experiment' which most directly probes the vacuum is the
observation of the cosmological constant. As we have argued there
is a striking difference between this observation and the quantum
effects which are usually taken to point to a substantial conception
of vacuum: Whereas the latter effects result from quantum field
theoretic considerations which might refer to the
constituents of the measurement arrangements rather than
properties of the vacuum, the observation of $\Lambda$ in e.g. the
solar system rests on purely classical measurement arrangements
which cannot possibly be held responsible for the observed value
of $\Lambda$ (if this observation is thought to refer to a vacuum
energy of quantum origin).

That the cosmological constant is observed to be zero or close to zero
could therefore suggest that there is no real vacuum energy of empty
space. At least, it should be scrutinized further if the standard
conception of the QFT vacuum involve unjustified extrapolations beyond
what is experimentally seen.  Apart from the QED and QCD contributions
to the vacuum energy, we have discussed how the Higgs mechanism in the
electroweak theory is believed to imply an enormous vacuum energy and
that larger vacuum energies in the context of grand unified theories
could have been the source of inflation.  It should be borne in mind,
however, that since, as yet, no Higgs particle has been found, and no
clear confirmation for inflation has been established, these
motivations for conceiving the vacuum energy as physical real remain
speculative.\\

\noindent
{\em 2. The existence of a link between $\Lambda$ and quantum vacuum energy}

\noindent
If one grants the reality of quantum vacuum energy one must also
assume an inter-theoretical link between QFT and GR, specifically
that quantum vacuum energy is a source of curvature in GR, in order
to establish the cosmological constant problem.  But, as we have seen,
it can be doubted if the cosmological constant problem is well-defined
-- given our insufficient understanding of this link between GR and
QFT. No doubt that both GR and QFT have had spectacular successes in
respectively the macroscopic and microscopic domain, but the gap
between these domains remains un-bridged.\footnote{See also
\cite{zinkernagel99} where it is discussed how the unity between
cosmology and particle physics is not as observationally well-founded
as is often assumed.}  In fact, as we have discussed, there are no
experimental indications of a relationship between QFT and GR, so this
relationship is based solely on theoretical expectations.

As concerns the theoretical expectations for a QFT-GR
relationship, we have discussed some of the technical and
conceptual difficulties related to the semi-classical approach in
which quantum fields are treated in a classical curved spacetime
background. These difficulties involve the problem of calculating
a finite value for $<\hat{T}_{\mu\nu}>$ in a curved background,
the problem of how to take the back-reaction effects of
$<\hat{T}_{\mu\nu}>$ on the metric into account, and the
difficulty of even defining a vacuum state in a curved background.
Nevertheless, we suggested that such difficulties are somehow
weakened by the fact that the cosmological constant
observationally is close to zero. In our view, this means that the
effects of treating QFT in the nearly flat spacetime background of
our Universe, instead of a strict Minkowski spacetime, are
probably small so that, for instance, a reasonable approximate
notion of the vacuum state can be found. While this observational
point may help to make the cosmological constant problem
well-defined in a semi-classical context, it does not, of course,
justify the existence of the link between the quantum vacuum
energy and $\Lambda$ (just because we in a consistent way can
define something does not imply that it exists). 
It would of course be an overwhelming surprise if physical
real vacuum energy did not gravitate since this would point to 
a serious misunderstanding in the standard expectations for
the connection between quantum field theory and the theory
of gravitation. The existence of such a connection and, more
generally, the establishment of a common framework for the 
description of all the fundamental forces, are major incentives
for modern theoretical physics. The experimental pillars on which
such incentives are based should, however, continuously be
illuminated.

\ \\
{\em Acknowledgements}

\noindent
It is a pleasure to thank Holger Bech Nielsen and Benny Lautrup
for discussions on the subjects of the present manuscript. We
would also like to thank Jes\' us Moster{\'\i}n for comments and
discussion on earlier drafts of this paper. Finally, we thank
the Niels Bohr Archive in Copenhagen for kindly providing
guided access to Niels Bohr's scientific correspondence.

\end{document}